\documentclass[
reprint,onecolumn,
amsmath,amssymb,
aps,
pra,
]{revtex4-2}

\usepackage{graphicx}
\usepackage{dcolumn}
\usepackage{bm}
\usepackage{float}
\usepackage{enumerate}
\usepackage{subfigure}
\usepackage[cmyk]{xcolor}
\usepackage{multirow}
\usepackage{makecell}
\usepackage{color}
\usepackage{amsmath,amssymb}
\usepackage{hyperref}
\hypersetup{colorlinks=true,citecolor=blue,linkcolor=blue,filecolor=blue,urlcolor=blue,breaklinks=true}

\newcommand{\nc}{\newcommand}
\nc{\rnote}[1]{\marginpar{\small\raggedright #1}}
\nc{\txtr}[1]{\textcolor{red}{#1}}
\nc{\txtb}[1]{\textcolor[RGB]{69,208,248}{#1}}
\nc{\txtg}[1]{\textcolor[RGB]{56,168,68}{#1}}
\nc{\txtbr}[1]{\textcolor{brown}{#1}}

\nc{\rr}[1]{\rnote{\txtr{#1}}}
\nc{\rg}[1]{\rnote{\txtg{#1}}}
\nc{\rb}[1]{\rnote{\txtb{#1}}}

\nc{\bra}[1]{\langle#1|}
\nc{\ket}[1]{|#1\rangle}
\nc{\ketbra}[2]{|#1\rangle\!\langle#2|}
\nc{\braket}[2]{\langle#1|#2\rangle}
\nc{\Img}{\operatorname{Im}}
\nc{\tr}{\operatorname{tr}}

\begin{document}


\title{The Top Manifold Connectedness of Quantum Control Landscapes}

\author{Yidian Fan}
\affiliation{%
Department of Automation, Tsinghua University, Beijing, 100084, China}%
\author{Tak-San Ho}
\affiliation{%
Department of Chemistry, Princeton University, Princeton, NJ 08544, USA}%
\author{Gaurav Bhole}
\affiliation{%
Department of Chemistry, Princeton University, Princeton, NJ 08544, USA}%
\author{Re-Bing Wu}
\thanks{E-mail: rbwu@tsinghua.edu.cn}
\affiliation{%
Department of Automation, Tsinghua University, Beijing, 100084, China}%
\author{Herschel Rabitz}
\thanks{E-mail: hrabitz@princeton.edu}
\affiliation{%
Department of Chemistry, Princeton University, Princeton, NJ 08544, USA}%

\date{\today}

\begin{abstract}
The control of quantum systems has been proven to possess trap-free optimization landscapes under the satisfaction of proper assumptions. However, many details of the landscape geometry and their influence on search efficiency still need to be fully understood. This paper numerically explores the path-connectedness of globally optimal control solutions forming the top manifold of the landscape. We randomly sample a plurality of optimal controls in the top manifold to assess the existence of a continuous path at the top of the landscape that connects two arbitrary optimal solutions. It is shown that for different quantum control objectives including state-to-state transition probabilities, observable expectation values and unitary transformations, such a continuous path can be readily found, implying that these top manifolds are fundamentally path-connected. The significance of the latter conjecture lies in seeking locations in the top manifold where an ancillary objective can also be optimized while maintaining the full optimality of the original objective that defined the landscape.
\end{abstract}

\maketitle

\section{Introduction}
\label{sec:I}
The past two decades have witnessed remarkable achievements in optimal control experiments of quantum phenomena driven by electromagnetic fields~\cite{rabitz2000whither,weiner2000femtosecond,wollenhaupt2005quantum}, including the manipulation of molecular states in chemical reactions~\cite{levis2002closing,bartels2002nonresonant,assion1998control,herek2006coherent}, and the synthesis of high-fidelity quantum gates for quantum computing~\cite{ramakrishna1996relation,nielsen2010quantum,krantz2019quantum}. These experiments collectively support that finding a high-quality control field is relatively easy. Optimal control simulations also show that even simple gradient-based optimization algorithms can almost always yield globally optimal solutions~\cite{ho2006effective}. Upon satisfaction of proper assumptions, these findings can be understood based on the topology of the underlying quantum control landscape (QCL), depicting the optimization objective $J$ as a functional of the control field $E(t)$. The most familiar control objectives include the transition probability between distinct quantum states, the expectation value of a physical observable, or a suitable norm of the fidelity between the evolution operator and a target unitary transformation. Under the proper assumptions~\cite{rabitz2006topology,rabitz2006optimal,rabitz2005landscape}, one can prove that no local extrema exist in these trap-free landscapes, which explains the ubiquitous successes in quantum optimal control.

\begin{figure}
	\centering
	\includegraphics[width=1\columnwidth]{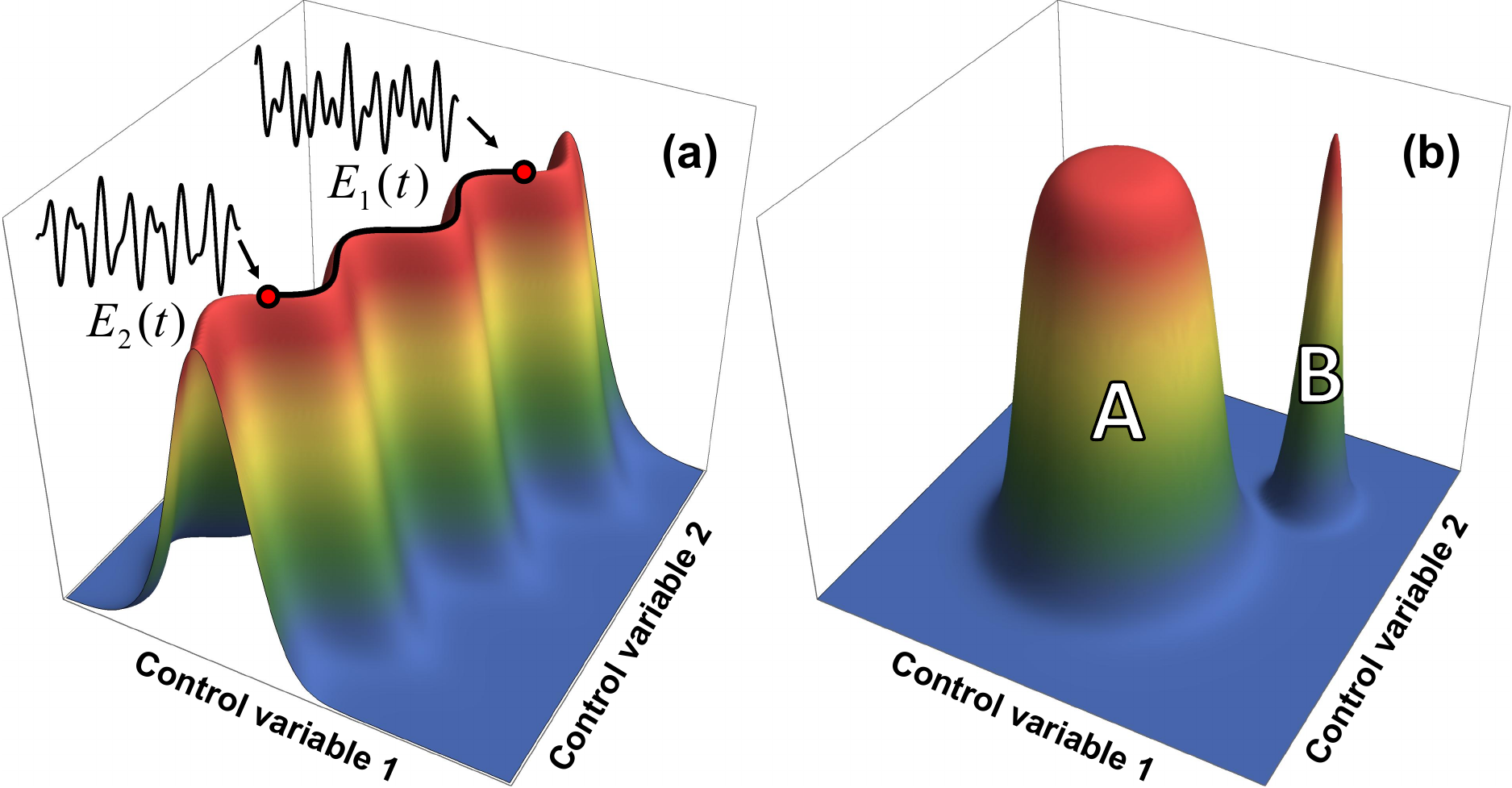}
	\caption{Schematics of two examples of top manifolds for QCLs having differing connectedness properties. The landscape is illustrated as the objective function $J$ over two control variables (usually hundreds of variables in practice) (a) A connected top manifold formed by optimal controls. Two optimal control fields, $E_1(t)$ and $E_2(t)$ (depicted by the red dots), can be connected by a continuous path (black line) on the top. (b) A disconnected top manifold composed of two separate regions in which the flatter region of $A$ implies higher robustness against control noise than that of a local point at the top of $B$.}
	\label{fig:qcl_schematics}
\end{figure}

Beyond finding a viable control, the existence of diverse optimal controls is also important in any particular application~\cite{demiralp1993optimally}. The top manifold that is spanned by optimal controls corresponds to the highest level set of the control landscape. The D-MORPH algorithm is amenable to identifying continuous regions of optimal controls by exploring the null space of the local Hessian over the top manifold~\textcolor{green}{\cite{beltrani2011exploring}}. For example, one may envisage a solution with some desired property, including low fluence~\cite{rothman2005observable}, high robustness~\cite{beltrani2011exploring},  smoothness~\cite{larocca2020exploiting}, etc. To this end, it is fundamental to understand the possibility of path-connectedness between the regions of optimal controls forming the highest level set of the control landscape. Of special interest is whether an optimal solution can be continuously transformed to any other solution while remaining on the top. There may exist particular regions that can be easily reachable starting from another region, if path-connectedness is satisfied, as sketched in Fig.~\ref{fig:qcl_schematics}(a). Moreover, disconnected regions may have distinct shapes and properties, for example, the solutions on the broad, flat region of $A$ in Fig.~\ref{fig:qcl_schematics}(b) are robust against control noises. In contrast, the solutions on the narrow, pointed region of $B$ are susceptible to disturbances~\cite{rabitz2004quantum,hocker2014characterization}. Controls between path disconnected regions may be reachable using a stochastic search algorithm as in seeking optimal controls that go from the top of region A to B in Fig.~\ref{fig:qcl_schematics}(b). In contrast, this paper strictly uses diffeomorphic algorithms (i.e., based on the D-MORPH method~\cite{rothman2005observable,rothman2006exploring,beltrani2010level}) in testing for top manifold connectedness to assess its scope, particularly for controls far from each other in the top manifold.

Since the top manifold is embedded in a high-dimensional control space, its connectedness is complicated to analyze. Numerical simulations have shown that the top manifold usually splits into disconnected regions when the accessible control variables are severely constrained~\cite{beltrani2011exploring,larocca2020exploiting,moore2012exploring,tibbetts2015constrained}. The conjecture of top manifold path-connectedness was rigorously proven for two level systems in the extended control space incorporating the control pulse duration $T$ as an additional variable~\cite{dominy2012dynamic} constituting a strongly controllable situation. However, no conclusion has been drawn on the connectedness of the top manifold for general controllable systems with a fixed pulse duration $T$ regardless of the control resources.

Interestingly, the QCL shares considerable similarities with the loss landscape of deep neural networks. It has been shown theoretically under proper assumptions that the local descent method can locate the global minimum for a wide range of nonlinear neural networks~\cite{sun2020global}. The loss landscape possesses multiple connected optima (referred to as mode connectivity), which may explain why deep neural networks are highly generalizable~\cite{draxler2018essentially, garipov2018loss, izmailov2019averaging} and robust against noise~\cite{kuditipudi2019explaining, Zhao2020BridgingMC}. The mode connectivity can be proven rigorously if the network is sufficiently wide~\cite{nguyen2019connected,nguyen2021note}, or if the optimal network solutions satisfy particular stability conditions~\cite{kuditipudi2019explaining,nguyen2021when}. Recent studies even suggest that simple linear interpolations can connect the solutions (up to proper permutations~\cite{HECHTNIELSEN1990129}) found by stochastic gradient descent \cite{entezari2022the,ainsworth2023git,Ferbach2024}. In addition to the theoretical analysis, the implementation of the string method has numerically shown low-loss landscape path-connectedness for some optimal network solutions~\cite{draxler2018essentially}. The string method is an adapted version of the nudged elastic band method, which was originally utilized to locate a minimum energy path connecting two metastable states corresponding to the minima of a potential energy surface~\cite{henkelman2000climbing,weinan2002string}. This method was also applied to parameterized quantum circuits used to construct quantum neural networks~\cite{hamilton2022mode}.

In this paper, we will apply the string method and D-MORPH algorithm to perform large-scale numerical simulations for testing the path-connectedness of the top manifold. Section~\ref{sec:II} provides the background and the formulation of the problem. Section~\ref{sec:III} describes the numerical methods. Section~\ref{sec:IV} presents optimal control simulations of path-connectedness on a representative four-level control system, which reveals the structure of the top manifold qualitatively and quantitatively. Section~\ref{sec:V} generalizes the findings to vast regions of the top manifold and addresses other classes of quantum systems. Finally, conclusions are drawn in Sec.~\ref{sec:VI}.

\section{Problem Formulation}
\label{sec:II}

\subsection{Background on common quantum control objectives}
\label{subsec:II-A}
Consider a closed $N$-level quantum system, whose state (density matrix) evolves as $\rho(t)=U(t)\rho_0 U^{\dagger}(t)$, where $\rho_0$ is the initial state at time $t=0$. Here, $U(t)$ is the system's unitary propagator, which is within the $N$-dimensional unitary group $\operatorname{U}(N)$ (or the special unitary group $\operatorname{SU}(N)$, as appropriate) and governed by the following Schrödinger equation (in units where $\hbar=1$):
\begin{equation}\label{eq:schrodinger}
	\frac{dU(t)}{dt}=-i\left[H_0-\mu E(t)\right]U(t),~~U(0)=\mathbb{I}_N,
\end{equation}
where $H_0$ is the field-free Hamiltonian and $\mu$ is the dipole moment operator. $E(t)$ is a real-valued control field over the time interval $[0,T]$, where $T$ is the duration of control pulse, and $\mathbb{I}_N$ is the $N$-dimensional identity matrix. The primary goal of quantum control is to maximize some objective $J$ as a functional of the control field $E(t)$. In this paper, we focus on the following three control landscapes that can be expressed as a function of the unitary transformation $U(T)\in \operatorname{U}(N)$ at the final time $T$.

(I) The \emph{state transition landscape} (STL), which aims at maximizing the transition probability between the initial and final pure states $|i\rangle$ and $|f\rangle$ at time $T$:
\begin{equation}\label{eq:j_stl}
	J_P(U(T))=|\langle f|U(T)|i\rangle|^2.
\end{equation}

(II) The \emph{observable control landscape} (OCL), which aims at maximizing the expectation value of a particular observable $\theta$:
\begin{equation}\label{eq:j_ocl}
	J_{\theta}(U(T))=\operatorname{tr}[\rho(T) \theta]=\operatorname{tr}[U(T)\rho U(T)^{\dagger} \theta].
\end{equation}

(III) The \emph{unitary transformation landscape} (UTL), which aims at maximizing the fidelity between $U(T)$ and the target unitary transformation $W$:
\begin{equation}\label{eq:j_utl}
	J_W(U(T))=\frac{1}{2}+\frac{1}{2N}\operatorname{Re} \operatorname{tr}\left(W^{\dagger} U(T)\right),
\end{equation}
where $W\in \operatorname{U}(N)$ is the target unitary gate that operates on quantum states. The target \eqref{eq:j_utl} is equivalent to $J_W=1-\frac{1}{4N}\|W-U(T)\|_{\rm F}^2$, where $\|\cdot\|_{\rm F}$ is the Frobenius norm~\cite{tibbetts2012exploring}. Here, the value of $J_W$ is normalized to the range $[0,1]$ and treated as the fidelity to make it consistently operative with the objectives \eqref{eq:j_stl} and \eqref{eq:j_ocl} (as shown in Sec.~\ref{sec:IV}). 

These landscapes are composed of two mappings: $\mathbb{K} \overset{U_T}{\longrightarrow} \operatorname{U}(N)\longrightarrow \mathbb{R}$, where $\mathbb{K}$ is the set of admissible controls. Here, the endpoint map 
$U_T:\mathbb{K}\rightarrow \operatorname{U}(N)$ maps a control field to the corresponding final unitary transformation $U(T)$ defined by Eq.~(\ref{eq:schrodinger}). The original objective $J\left[E(t)\right]$ defined on the control space $\mathbb{K}$ is denoted by the dynamical landscape, while $J[U(T)]$ defined on $\operatorname{U}(N)$ is called the kinematic landscape. In the dynamical landscape, a control field $E(t)$ is called a critical point if the first-order derivative of the objective $J$ vanishes, i.e.,
\begin{equation}\label{eq:j_gradient}
	\frac{\delta J}{\delta E(t)}=\left\langle\nabla J\left[U(T)\right], \frac{\delta U(T)}{\delta E(t)}\right\rangle=0,~~\forall t\in [0,T],
\end{equation}
where $\nabla J\left[U(T)\right]$ is the gradient of $J$ with respect to $U(T)$, $\frac{\delta U(T)}{\delta E(t)}$ is the Frechet derivative (or Jacobian) of $U(T)$ at $E(t)$, and $\langle \cdot,\cdot\rangle$ is the Hilbert-Schmidt inner product. 

Equation~(\ref{eq:j_gradient}) reveals the connection between the dynamical and kinematic landscapes, and one can prove that $E(t)$ is a critical point if and only if 
\begin{equation}\label{eq:j_critical}
	\nabla J\left[U(T)\right]=0, 
\end{equation}
when the following three conditions are satisfied~\cite{rabitz2005landscape,rabitz2006optimal,brif2010control}: (i) the closed quantum system (\ref{eq:schrodinger}) is controllable~\cite{ramakrishna1995controllability}; (ii) the mapping $U_T$ from the control field $E(t)$ to the unitary transformation $U(T)$ is locally surjective, i.e., the Jacobian $\frac{\delta U(T)}{\delta E(t)}$ has full rank; (iii) the control is unconstrained. Condition (ii) implies that the dynamical landscape has no traps (i.e., local optimum) if and only if the kinematic landscape does not have traps \cite{wu2008control}, which holds for all the three cases given by Eqs.~(\ref{eq:j_stl})-(\ref{eq:j_utl})~\cite{dominy2012dynamic, hsieh2008optimal}. It should be noted that condition (ii) may be violated at a so-called singular control, where the Jacobian is rank-deficient~\cite{wu2012singularities}. Nevertheless, numerical simulations suggest that singular controls have a minor impact on optimization~\cite{riviello2014searching}. Hence, the landscapes can be generally treated as trap-free if conditions (i) and (iii) are satisfied.

\subsection{Path-connectedness of quantum control landscape top manifold}
A path between optimal controls $x$ and $y$ in the top manifold $\mathcal{M}$ is defined as the image of a continuous function $f:[0,1]\rightarrow \mathcal{M}$ such that $f(0)=x$ and $f(1)=y$. The top manifold is path-connected if every pair of points in $\mathcal{M}$ can be connected by a path in $\mathcal{M}$~\cite{munkres2000topology}. It is not difficult to prove that the top manifolds of the kinematic landscapes in Eqs.~(\ref{eq:j_stl})-(\ref{eq:j_utl}) are path-connected~\cite{hsieh2008optimal,dominy2012dynamic}, but the analysis of the corresponding dynamical landscapes is much more difficult due to the high complexity of the endpoint mapping $U_T:\mathbb{K}\rightarrow \operatorname{U}(N)$. 

As remarked before, the top manifold is path-connected when the system is strongly controllable, and the control pulse duration $T$ is free to vary~\cite{dominy2012dynamic}. One also can find an uncontrollable system whose top manifold is connected in the kinematic picture but disconnected in the dynamical picture. For example, the following two-level system with
\begin{equation}
	H_0=\begin{pmatrix} 
		0 & 0 \\
		0 & 0 \\ 
	\end{pmatrix},  
	\mu=\begin{pmatrix} 
		0 & 1 \\
		1 & 0 \\ 
	\end{pmatrix}
\end{equation}
is uncontrollable~\cite{d2021introduction}. The unitary transformation $U(t)$ can be analytically solved according to Eq.~(\ref{eq:schrodinger}):
\begin{equation}
	U(t)=\begin{pmatrix} 
		\cos\varphi(t) & i\sin\varphi(t) \\
		i\sin\varphi(t) & \cos\varphi(t) \\ 
	\end{pmatrix},
\end{equation}
in which $\varphi(t)=\int_0^t E(t')dt'$. For the UTL with $W=\begin{pmatrix} 
	0 & i \\
	i & 0 \\ 
\end{pmatrix},$ 
$J_W$ reaches the maximum in the top manifold consisting of an isolated point $U=W$ in the kinematic landscape, so path-connectedness apparently holds. In the dynamical picture, the top manifold consists of disconnected subsets corresponding to $\varphi(T)=2k\pi+\frac{\pi}{2}$, $k\in \mathbb{Z}$. 

Therefore, throughout this paper we will only consider the controllable quantum systems with fixed control pulse duration $T$ and investigate whether their top manifolds are path-connected.

\section{Numerical methods for exploring the top manifold}
\label{sec:III}
Since the explicit structure of the top manifold has not been resolved analytically, we will probe the degree of path-connectedness and encountered control space `features' by numerically seeking continuous paths in the top manifold between a plethora of sampled optimal controls. The successful identification of such paths would serve as evidence that the top manifold is path-connected. This numerical investigation also provides a hint for future theoretical analysis on this topic discussed in the Conclusion Section~\ref{sec:VI}. The following subsections will describe the numerical methods for performing the simulations to test path-connectedness.

\subsection{Sampling optimal controls via the gradient ascent algorithm}
To test path-connectedness, we first need to generate a plurality of optimal fields in the top manifold. By leveraging the trap-free property of landscapes, the local gradient-based method can efficiently locate optimal control fields. The algorithm operates by introducing a variable $s\geq 0$ labelling the control field $E(s,t)$ and its objective $J(s)$ upon the landscape climb. The differential change of the $J(s)$ value is given by the chain rule as
\begin{equation}\label{eq:chain_rule}
	\frac{dJ(s)}{ds}=\int_0^T \frac{\delta J}{\delta E(s,t)}\frac{\partial E(s,t)}{\partial s} dt,~~s\geq 0.
\end{equation}
The gradient flow is described by
\begin{equation}\label{eq:gradient_flow}
	\frac{\partial E(s,t)}{\partial s}=\alpha(s) \frac{\delta J}{\delta E(s,t)},
\end{equation}
where $\alpha(s)>0$ controls the rate of gradient ascent. The gradient flow ensures that
\begin{equation}
	\frac{dJ(s)}{ds}=\alpha(s)\int_0^T \left(\frac{\delta J}{\delta E(s,t)}\right)^2 dt\geq 0
\end{equation}	
and brings the initial guess $E(0,t)$ up to a point $E(s,t)$ very close to the top manifold when $s$ is sufficiently large. The functional derivative on the right-hand side of Eq.~(\ref{eq:gradient_flow}) has the following forms with respect to the three considered landscapes~\cite{riviello2017searching, tibbetts2012exploring}:
\begin{equation}
	\begin{aligned}
		\frac{\delta J_P}{\delta E(t)}&=2 \Img \tr \left[U^{\dagger}(T) |f\rangle \langle f| U(T) |i\rangle \langle i| \mu(t)\right],\\
		\frac{\delta J_{\theta}}{\delta E(t)}&=2 \Img \tr\left[U^{\dagger}(T) \theta U(T) \rho_0 \mu(t)\right],\\
		\frac{\delta J_W}{\delta E(t)}&=-\frac{1}{2N} \Img \tr\left[W^{\dagger} U(T)\mu(t)\right],
	\end{aligned}
\end{equation}
where $\mu(t)=U^{\dagger}(t)\mu U(t)$. 

In our simulations, the optimal controls are sampled by numerically solving Eq.~(\ref{eq:gradient_flow}) with the MATLAB routine \verb+ode45+ from random initial trial fields. Equations~(\ref{eq:gradient_flow}) and (\ref{eq:schrodinger}) are sequentially solved in the process. The optimization terminates when $E(s, t)$ is sufficiently close to the top, i.e., when $J\left[E(s, t)\right]\geq J_{\max}-\varepsilon$, where $J_{\max}$ is the maximum objective value and $\varepsilon$ is a small positive number (tolerance) close to zero.  Hereafter, we will specify a small value of $\varepsilon$ (later the sensitivity to its value will be tested) and define $J_{\max}-\varepsilon$ as the practical "maximum" top manifold of the landscape, while efficiently moving over this high level set for numerical reasons, as will be explained later. Unless otherwise specified, $J_{\max} - \varepsilon$ will be referred to as the top manifold.

\subsection{Path-connectedness tests via the string method}
\begin{figure}[htbp]
	\centering
	\includegraphics[width=\columnwidth]{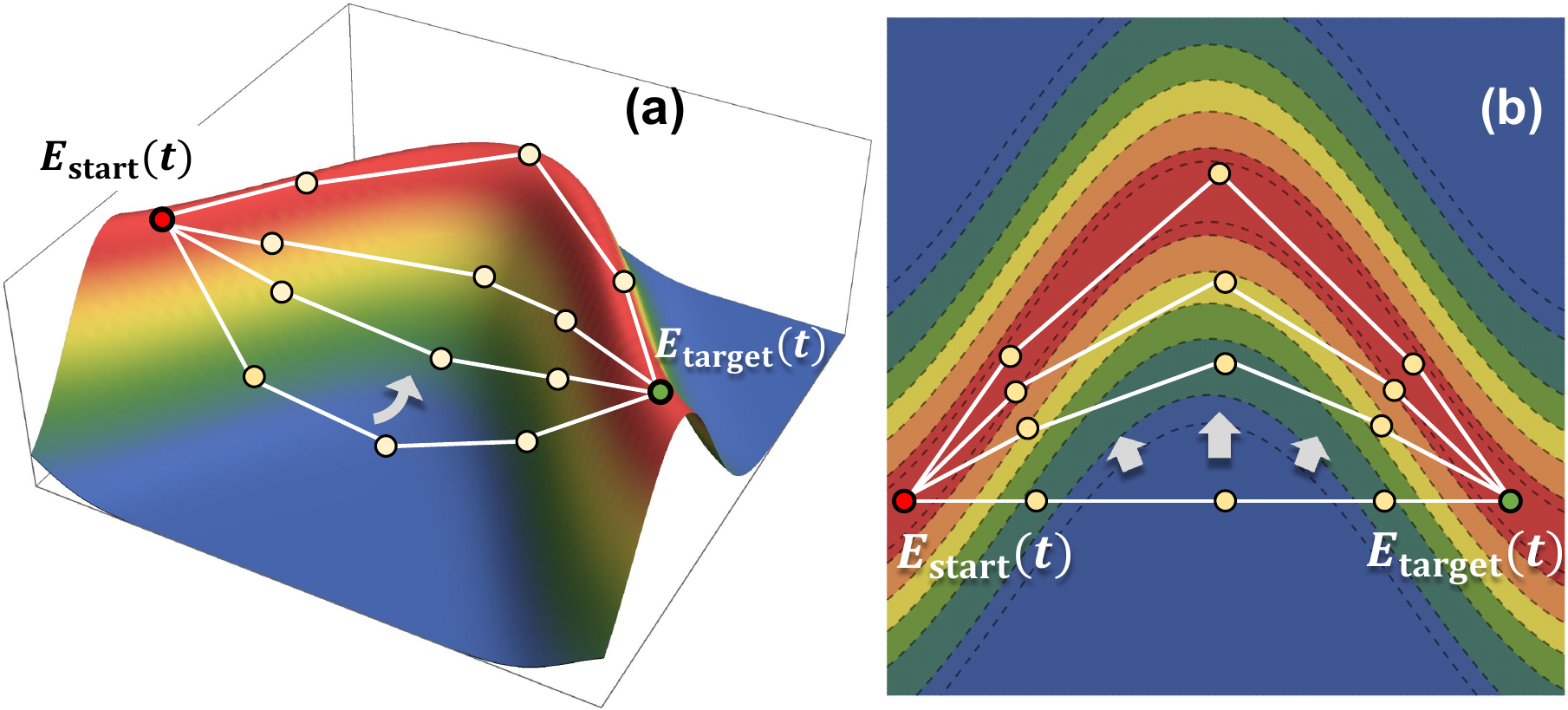}
	\caption{Schematics of successful use of the string method. (a) A portion of the QCL specified in Fig.~\ref{fig:qcl_schematics}(a). (b) The two-dimensional contour plot of the landscape. The string (white line) is first initialized as a straight line sampled by interpolated fields (interior yellow circles) between two given optimal controls $E_{\rm {start}}(t)$ and $E_{\rm {target}}(t)$ (interior red and green circles, respectively). Then, it gradually evolves following the gradient flows (gray arrows) and associated strings until reaching the top manifold.}
	\label{fig:string_schematics}
\end{figure}

Inspired by the algorithm studying the landscape of neural networks~\cite{draxler2018essentially}, we combine the simplified string method~\cite{weinan2007simplified} and the automated nudged elastic band method~\cite{kolsbjerg2016automated} aiming to discover a continuous path in the top manifold of quantum control landscapes that connects two arbitrary optimal controls $E_{\rm start}(t)$ and $E_{\rm target}(t)$. As illustrated in Fig.~\ref{fig:string_schematics}, the basic concept is to initialize a straight string sampled by interpolated fields between the two controls. The fields along the initial string are generally not in the top manifold except for its two ends. Thus, we push up the entire discretely sampled set of fields along the string through the gradient flow, aiming for every intermediate control field to be sufficiently close to the top manifold. If this can be successfully done, a path consisting of many intermediate fields will be obtained in the top manifold connecting the two fields, $E_{\rm start}(t)$ and $E_{\rm target}(t)$. Otherwise, the failure of this method will lead to what we call a broken string, indicating that the two latter control fields cannot be connected by a continuous path in the top manifold. The detailed algorithm is described in Appendix~\ref{app:string_method}. The reason for moving in a high level set $J_{\max}-\varepsilon$ is to avoid dealing with the Hessian at the actual top of the landscape $J_{\max}$~\cite{beltrani2011exploring}; this issue is a practical matter as the Hessian is expensive to evaluate while the gradient $\frac{\delta J}{\delta E(s,t)}$ is operative at a high level set with due attention to the stability of numerical integration. The effects of varying $\varepsilon$ will be shown later as insignificant, provided that it is sufficiently small.

\subsection{Path-connectedness tests via the D-MORPH algorithm}\label{subsec:iii-c}
Although the D-MORPH method in Eq.~(\ref{eq:gradient_flow}) is often referred to as an effective algorithm for climbing the landscape, it was initially formulated for moving in a level set including the top manifold~\cite{rothman2006exploring,beltrani2007photonic,beltrani2010level}. In the latter context, the D-MORPH method is a homotopy algorithm for exploring the landscape level set. In this context, we will use the diffeomorphic variable $s\geq 0$ to characterize the movement over the near top manifold. Given a starting point $E(0,t)$ in the top manifold, the D-MORPH algorithm is designed to generate a continuous trajectory $E(s,t)$, along which the objective function $J\left[E(s,t)\right]$ stays constant at a high level set $J_{\max}-\varepsilon$, i.e., $\frac{dJ(s)}{ds}=0$. We define a projector $\hat{P}$ that acts on an arbitrary function $f(s,t)$ and projects it into the space specified by the gradient $\frac{\delta J}{\delta E(s,t)}$ such that
\begin{equation}\label{eq:projector_arbitrary}
	\hat{P}\cdot f(s,t)=\frac{\int_0^T\frac{\delta J}{\delta E(s,t')}f(s,t')dt'}{\int_0^T \left(\frac{\delta J}{\delta E(s,t')}\right)^2 dt'}\frac{\delta J}{\delta E(s,t)}.
\end{equation}
The differential change $\frac{\partial E(s,t)}{\partial s}$ aiming to maintain $J_{\rm max} - \varepsilon$ as constant given within the D-MORPH algorithm has the form
\begin{equation}\label{eq:dmorph_grad}
	\frac{\partial E(s,t)}{\partial s} =[\mathbf{1}-\hat{P}]\cdot f(s,t),
\end{equation}
where $\mathbf{1}\cdot f(s,t)=f(s,t)$. Substituting Eq.~(\ref{eq:dmorph_grad}) into Eq.~(\ref{eq:chain_rule}) and utilizing Eq.~(\ref{eq:projector_arbitrary}) readily leads to the result that
\begin{align}\label{eq:move_levelset}
	\frac{d J(s)}{ds}& = 0.
\end{align}
Here, $f(s,t)$ guides the local motion of the trajectory on the very high top level set and $f(s,t)$ can be tailored according to a specified search objective. Finally, we remark that $[\mathbf{1}-\hat{P}]$ projects onto the near top manifold that is locally orthogonal to $\frac{\delta J}{\delta E(s,t)}$, $t\in [0,T]$.

\begin{figure}[htbp]
	\centering
	\includegraphics[width=1\columnwidth]{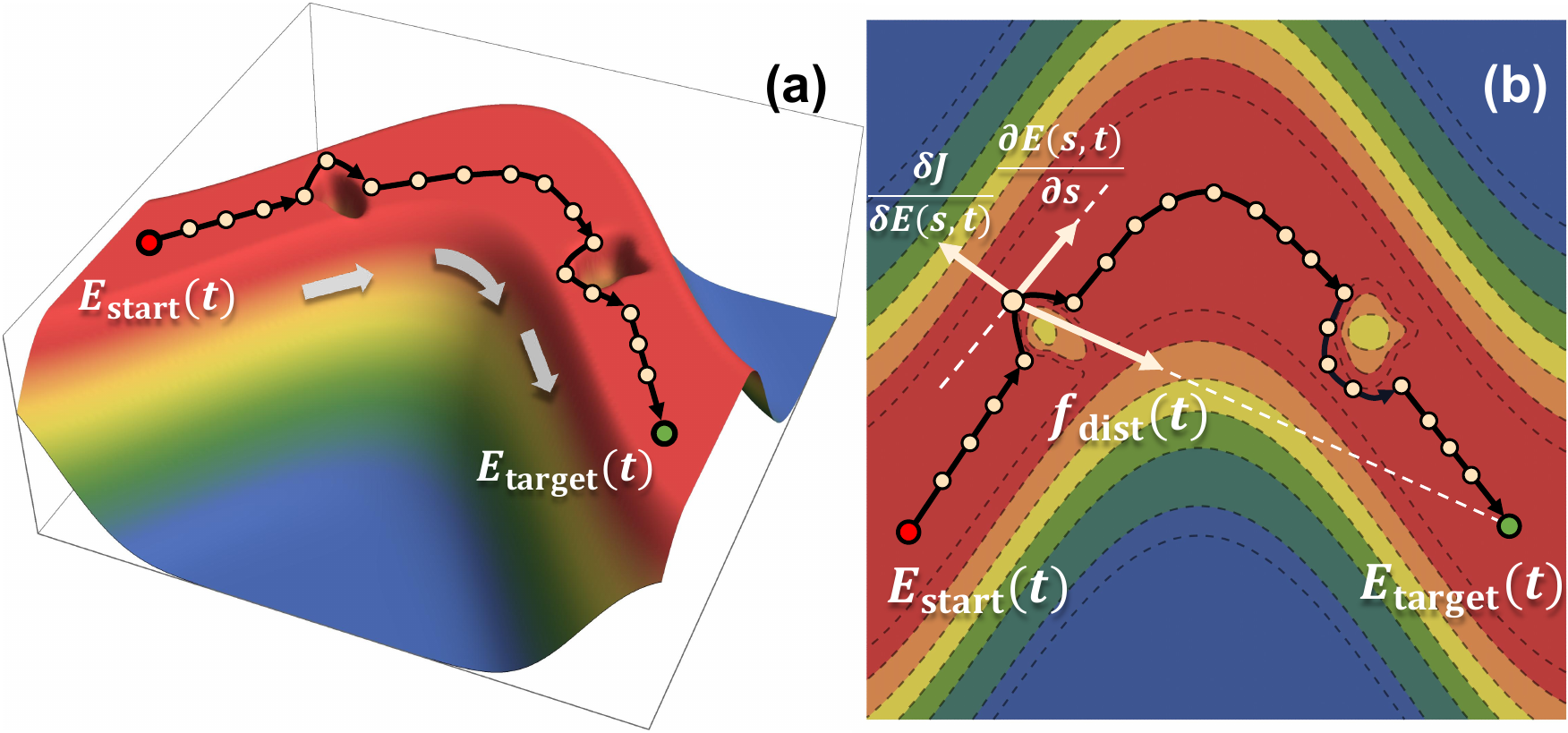}
	\caption{Schematics of the D-MORPH connecting algorithm. (a) A zoomed-in section of the top manifold specified in Fig.~\ref{fig:qcl_schematics}(a), highlighting the presence of two possible impeding features. (b) The two-dimensional contour plot of the landscape. The D-MORPH connecting algorithm explores the high-level set $J_{\max}-\varepsilon$. Starting from one optimal control field $E_{\rm {start}}(t)$ (interior red circles), the differential change $\frac{\partial E(s,t)}{\partial s}$ is determined as the projection of $f_{\rm {dist}}(s,t)$ onto the subspace orthogonal to the local gradient $\frac{\delta J}{\delta E(s,t)}$, as depicted in (b). Note that the D-MORPH method for this purpose has been built so that the trajectory~(black line), from $E_{\rm {start}}(t)$ towards $E_{\rm {target}}(t)$, remains on the top of the landscape. The numerical evidence shows that such trajectories almost always avoid getting trapped by a top manifold feature as schematically indicated, by moving on the edge in this figure. }
	\label{fig:schematics_dmorph}
\end{figure}

The D-MORPH algorithm can be implemented aiming to generate a continuous path from a starting field  $E(0,t)=E_{\rm {start}}(t)$ towards a target field $E_{\rm {target}}(t)$ by appropriately choosing the guiding function $f(s,t)$. Let $\Delta(s,t)=E_{\rm {target}}(t)-E(s,t)$, and 
\begin{equation}\label{eq:distance}
	D(s)=\|\Delta(s)\|^2_2
\end{equation}
be the square of the Euclidean distance from the current field to the target field, where $\|\cdot\|_2$ is the norm defined as $\|\Delta(s)\|_2=\left[\int_0^T \Delta^2(s,t)\: dt\right]^{\frac{1}{2}}$.  We desire for $D(s)$ to morph towards zero as $s$ proceeds. To this end, we choose
\begin{equation}\label{eq:guiding_func}
	f(s,t)=f_{\rm {dist}}(s,t):=\frac{\Delta(s,t)}{\|\Delta(s)\|_2}
\end{equation}
as the guiding function~\cite{tibbetts2015constrained}, which assures that
\begin{equation}\label{eq:monotonic_decrease}
	\frac{d D(s)}{d s} \leq 0
\end{equation}
while satisfying Eq.~(\ref{eq:move_levelset}). See Appendix~\ref{app:guiding_func} for details. Equation~(\ref{eq:monotonic_decrease}) implies that $D(s)$ will monotonically decrease towards zero if no hindering features in the top manifold are encountered when moving from $E_{\rm start}(t)$ towards $E_{\rm target}(t)$. Here, the normalization $\|f_{\rm {dist}}(s)\|_2=1$ ensures a steady pace of morphing while minimizing $\|\Delta(s)\|_2$. Figure~\ref{fig:schematics_dmorph} illustrates two schematics of the D-MORPH connecting algorithm, whose detailed procedure is summarized in Appendix~\ref{app:dmorph_connecting}. The algorithm is robust as Eq.~(\ref{eq:dmorph_grad}) continuously updates the guiding directions of the trajectory aiming to achieve $D(s)\to 0$. The local gradient information can allow the guiding trajectory to smoothly navigate around potential impeding features in the top manifold, as depicted by Fig.~\ref{fig:schematics_dmorph}. Although there is no guarantee that the D-MORPH algorithm can overcome every such feature, the extensive simulations in Sec.~\ref{subsec:IV-B} demonstrate that the exploration of the top manifold guided by Eq.~(\ref{eq:dmorph_grad}) with $f(s,t)$ defined in Eq.~(\ref{eq:guiding_func}) rarely halts. A full exploration of the nature of potential top manifold features awaits further research (see Conclusion Section~\ref{sec:VI}), but the indicator $R$ in Eq.~(\ref{eq:ratio_R}) below provides evidence that such features exist that must be navigated around for successfully finding top manifold connectivity between two arbitrary final fields. The success of the string method also reflects avoiding such features to prevent a broken string. A prior study of a two-level state-to-state top manifold found path-connectedness along with regions in control space outside of the top manifold (i.e., features), which were adjacent to each other (i.e., the edges of the feature form the boundary between top and non-top manifolds~\cite[Figure~9]{tibbetts2015constrained}.)

\subsection{Visualization of the connecting paths in the top manifold}
Once a continuous path is found connecting two arbitrary optimal controls in the top manifold, the structure of the path may contain a rich information about the nature of the high-dimensional top manifold. The degree of straightness of such a path $E(s,t)$ with $s\in [0,s_{\max}]$ is measured by the ratio, $R\geq 1$, of the path length $d_{\rm P}$ to the Euclidean distance $d_{\rm E}$ between two ends~\cite{nanduri2013exploring}, $E_{\rm start}(t)$ and $E_{\rm target}(t)$:
\begin{equation}\label{eq:ratio_R}
	R=\frac{d_{\rm P}}{d_{\rm E}}:=\frac{\int_0^{s_{\max }}\left[\int_0^T\left(\frac{\partial E(s, t)}{\partial s}\right)^2 d t\right]^{\frac{1}{2}} d s}{\left[\int_0^T\left(E\left(s_{\max }, t\right)-E(0, t)\right)^2 d t\right]^{\frac{1}{2}}}.
\end{equation}
In numerical simulations, the path is always sampled by discrete points $\{E_i(t):i=0,\cdots,N\}$ in control space, and thus the ratio $R$ can be approximately computed as $R=\frac{\sum_{i=0}^{N}\|E_i-E_{i-1}\|_2}{\|E_0-E_{N}\|_2}$. The path is nearly straight if the ratio $R\simeq 1$, while a high $R$ value suggests that the path is rather gnarled, indicating complex structure i.e., features in the top manifold. 

To get a reduced dimensional image of the actual control trajectory, we also use principal component analysis (PCA) proposed in Ref.~\cite{li2018visualizing} to project the path onto a three-dimensional space so that it can be visualized. This method is designed to find effective basis vectors that capture as much information about a high-dimensional path as possible in the most critical three-dimensional projection space (see Appendix~\ref{app:pca} for details). The percentage of the captured variation can be computed, and we will show that such three-dimensional projected trajectories provide an informative view of what the top manifold path looks like.

\section{Numerical tests of path-connectedness}
\label{sec:IV}
In this section, we apply the methods in Sec.~\ref{sec:III} to test path-connectedness of the top manifold with the following four-level quantum system:
\begin{equation}\label{eq:four_dim_sys_H0}
	H_0=\begin{pmatrix} -2.25 & 0 & 0 & 0 \\
		0 & -0.75 & 0 & 0 \\ 
		0 & 0 & 0.25 & 0\\
		0 & 0 & 0 & 2.75
	\end{pmatrix}, 
\end{equation} 
and
\begin{equation}\label{eq:four_dim_sys_dipole}
	\mu=\begin{pmatrix} 
		0 & 1 & 0.5 & 0.25 \\ 
		1 & 0 & 1 & 0.5 \\ 
		0.5 & 1 & 0 & 1\\ 
		0.25 & 0.5 & 1 & 0
	\end{pmatrix}.
\end{equation}
This system is fully controllable because $H_0$ and $\mu$ generate the Lie algebra ${\operatorname{SU}}(4)$~\cite{d2021introduction}. For the STL, we consider the population transfer from the state $|1\rangle$ to the state $|4\rangle$. For the OCL, the initial state and the physical observable are chosen as $\rho=\operatorname{diag}\{0.5,0.5,0,0\}$ and  $\theta=\operatorname{diag}\{0,0,1,1\}$, respectively. For the UTL, 
the target gate $W$ is specified as 
\begin{equation}
	W=e^{-\mathrm{i} \pi / 4}\left(
	\begin{array}{llll}
		1 & 0 & 0 & 0 \\
		0 & 1 & 0 & 0 \\
		0 & 0 & 0 & 1 \\
		0 & 0 & 1 & 0
	\end{array}\right).
\end{equation}

Each of the three landscapes has one top manifold and one bottom manifold corresponding to $J_{\max}=1$ and $J_{\min}=0$, respectively. In addition, there is one saddle manifold in the OCL and the UTL at the value $J=0.5$, while there are no saddle manifolds in the STL. 

\subsection{Distributions of optimal controls}
\label{IV-A}
We first sampled a large collection of optimal controls distributed in the top manifold of each objective using the gradient flow specified by Eq.~(\ref{eq:gradient_flow}). The initial trial fields, $E_{\rm init}(t)$, had the form:
\begin{equation}\label{eq:initial_fields}
	E_{\rm init}(t)=A\sum_{m=1}^{100}a_m \cos(\omega_m t+\phi_m),~~t\in [0,T].
\end{equation}
The amplitudes $\{a_m\}$ and the phase factors $\{\phi_m\}$ were randomly drawn from uniform distributions over $\left[0,1\right]$ and $\left[0,2\pi\right]$, respectively. The frequencies $\{\omega_m\}$ were chosen from the uniform distribution over $[0.5,5.5]$ that covered all allowed transition frequencies in $H_0$. The factor $A$ ensured the fluence $F_0$ of $E_{\rm init}(t)$ was normalized, i.e., $F_0:=\| E_{\rm init}\|_2^2=1$. The total control pulse duration was set to $T=50$ such that $\omega_mT\gg 1$ for all $\omega_m$ and the control pulse covered at least three oscillation cycles of the lowest frequency. In the simulations, $E(t)$ was evenly divided into $L=500$ pieces, with duration $\Delta t=\frac{T}{L}$, so that the control field could be represented by a $L$-dimensional vector $\left[E_{\rm init}(\Delta t),\cdots,E_{\rm init}(L\Delta t)\right]^T$.

\begin{figure}[htbp]
	\centering
	\includegraphics[width=1\columnwidth]{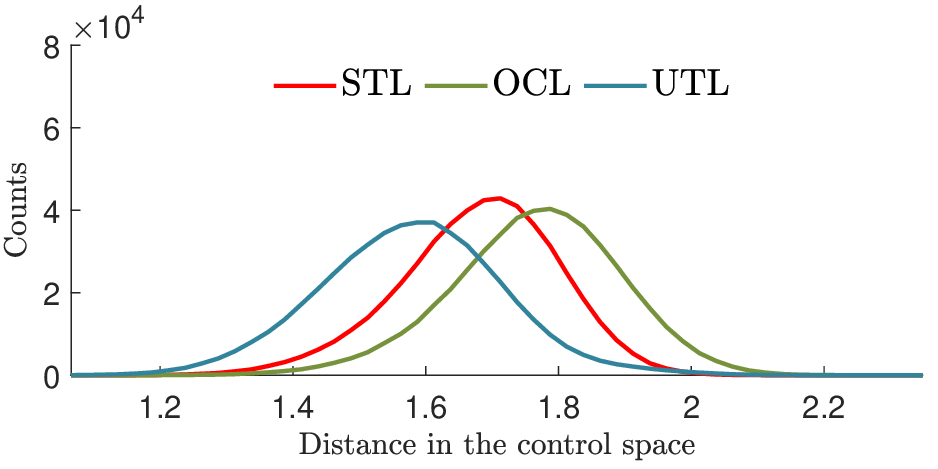}
	\caption{Distributions of the pairwise distances among optimal fields for STL (red), OCL (green), and UTL (blue).  Each distribution, which records  $\sim 5\times 10^5$ pairwise distances in total, is drawn from one thousand optimal controls in each landscape.}
	\label{fig:distribution_dist}
\end{figure}

One thousand trial controls $E_{i,{\rm init}}(t)$ were picked in each landscape according to Eq.~(\ref{eq:initial_fields}). By setting $E_i(s=0,t)= E_{i,{\rm init}}(t)$, each field was steered to a solution $E_{i,{\rm final}}(t):=E_i(s_{\max},t)$ in the top manifold by evolving Eq.~(\ref{eq:gradient_flow}) until $J[E_i(s_{\max},t)]\geq 0.999$ (i.e., $\varepsilon=0.001$). To characterize the distribution of the control fields, we calculated the pairwise distances among the optimal controls $d_{ij}=\| E_{i,{\rm final}}-E_{j,{\rm final}}\|_2$, where $1\leq i,j\leq 1000$, $i\neq j$. Their distributions are depicted in Fig.~\ref{fig:distribution_dist}. It can be seen that they all approximately obey a Gaussian shaped distribution, which is consistent with the finding in previous works~\cite{nanduri2013exploring,nanduri2015exploring}.

\subsection{Path-connectedness tests of the top manifold}
\label{subsec:IV-B}
Among the optimal controls obtained in Sec.~\ref{IV-A}, we randomly selected one thousand pairs $\{\left(E_{i,{\rm final}}(t), E_{j,{\rm final}}(t)\right):1\leq i,j \leq 1000, i\neq j\}$ to test their path-connectedness using the string method and the D-MORPH connecting algorithm. The ratio $R$ defined in Eq.~(\ref{eq:ratio_R}) calculated within the string method and the D-MORPH connecting algorithm were denoted by $R_{\rm st}$ and $R_{\rm dm}$, respectively. The top manifold tolerance in $J=J_{\max}-\varepsilon$ for both algorithms was set by choosing $\varepsilon=\varepsilon_{\rm st}=\varepsilon_{\rm dm}=0.001$.  

\begin{figure}
	\centering	\includegraphics[width=0.8\columnwidth]{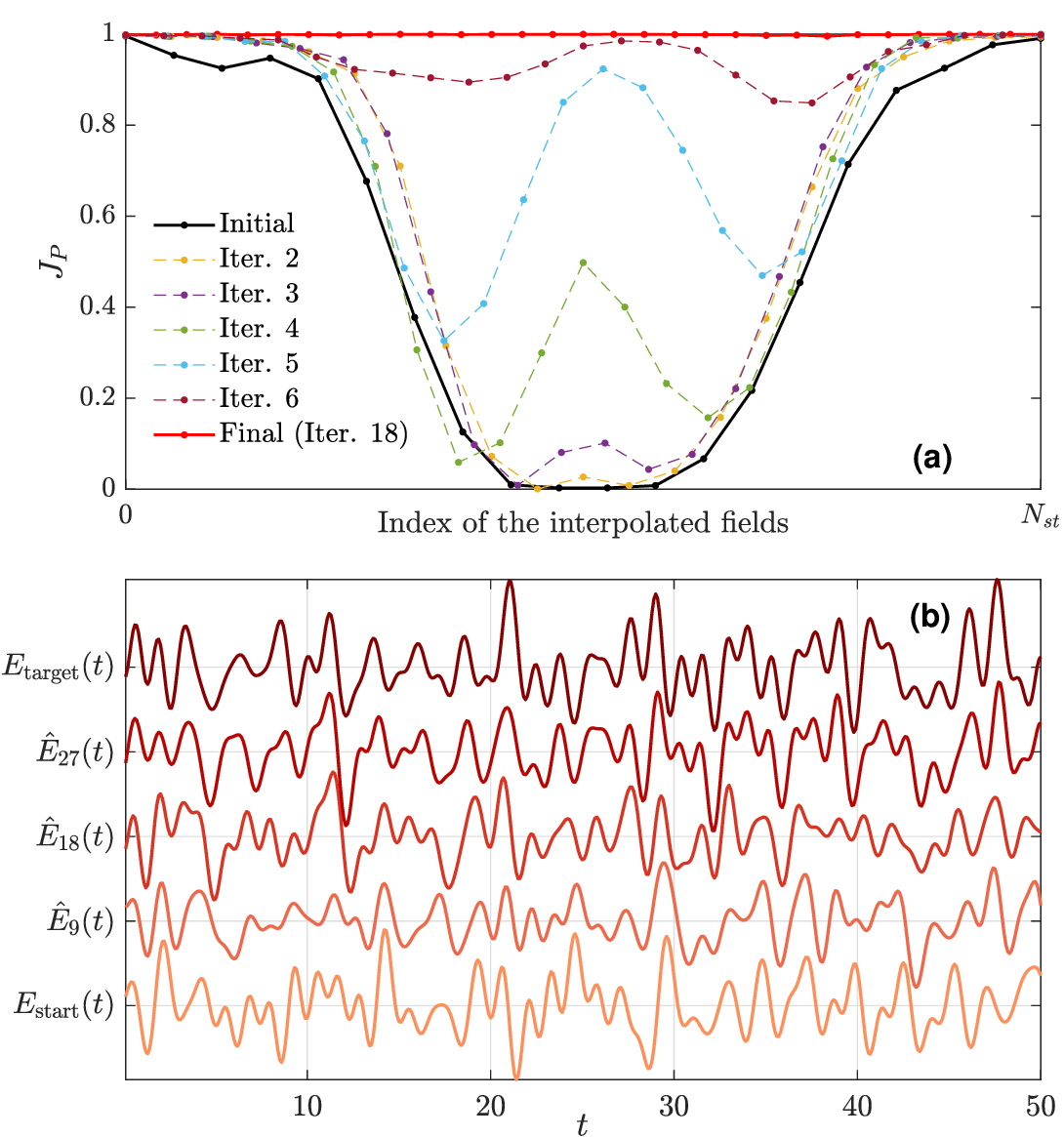}
	\caption{(a) The convergence process of the string method to assess landscape connectedness illustrated for a particular STL. The vertical axis is the objective function $J$. Each dot in the figure characterizes the objective function $J$ of one interpolated field along the string. Initialized as a straight line (black line) between the two optimal fields $(E_{\rm start}(t), E_{\rm target}(t))$ in the top manifold, the intermediate interpolated fields initially produce $J$ values far from the top manifold. The set of intermediate fields of the string is each subjected to an iterative climb along a gradient flow of the landscape, and the set of fields forming the string finally reaches the top manifold at the iteration $18$ (red line) for this case. (b) A succession plot of some of the interpolated fields $\{\hat{E}_i(t):i=0,1,\cdots, N_{\rm st}\}$ along the final converged string in the top manifold for this case, where $N_{\rm st}=34$. The interpolated fields (distinguished by color scale) demonstrate that the starting field $E_{\rm start}(t)$ (orange on the bottom of (b)) is continuously transformed to the final control $E_{\rm target}(t)$ (red on the top) along the string.}
    \label{fig:string_convergence}
\end{figure}

We first used the string method, aiming to locate continuous paths between the one thousand pairs. The simulation results show that all strings in the three landscapes converged successfully to curved paths in the top manifolds. Figure~\ref{fig:string_convergence}(a) displays the convergence process of one string in the STL scenario. The initial fields forming an interpolated straight string have an objective value $J$ that drops down and in this case even to the bottom of the landscape for some fields near the middle of the initial string as they are far from being in the top manifold. However, after being pushed by the gradient flows, the string (i.e., all of its $N_{st}$ member fields) finally reaches the top manifold with all interpolated fields yielding $J\geq 0.999$. Figure~\ref{fig:string_convergence}(b) plots the profiles of some of the interpolated fields along the \textit{final} converged string in the top manifold. Here we observe a smooth, but rather complex, transformation from one given optimal field to another along the final string. To test the interpolation inaccuracies, we randomly selected one hundred final strings in each landscape and evaluated the objective values of more densely sampled points along these strings. All evaluated fields yield $J\geq 0.998$, which verifies the robustness of the string method in identifying connected paths in the top manifold. This accuracy can be further increased by specifying stricter numerical tolerance $\varepsilon_{\rm st}$ (see Appendix \ref{app:string_method} for details). The behavior shown in Fig.~\ref{fig:string_convergence}(a) for the initial interpolated straight line in control space, especially the dip down from the top of the landscape, was found to some degree for all string method tests for the STL, OCL, and UTL. Thus, although path-connectedness was always found, this behavior indicates the encounter of features on the top of the landscape scattered about as mentioned above (i.e., moving along an interpolated \textit{straight} line from any two controls in the top manifold will likely encounter fields where $J$ significantly deviates from its top manifold value).

\begin{figure*}
	\centering
	\includegraphics[width=1\textwidth]{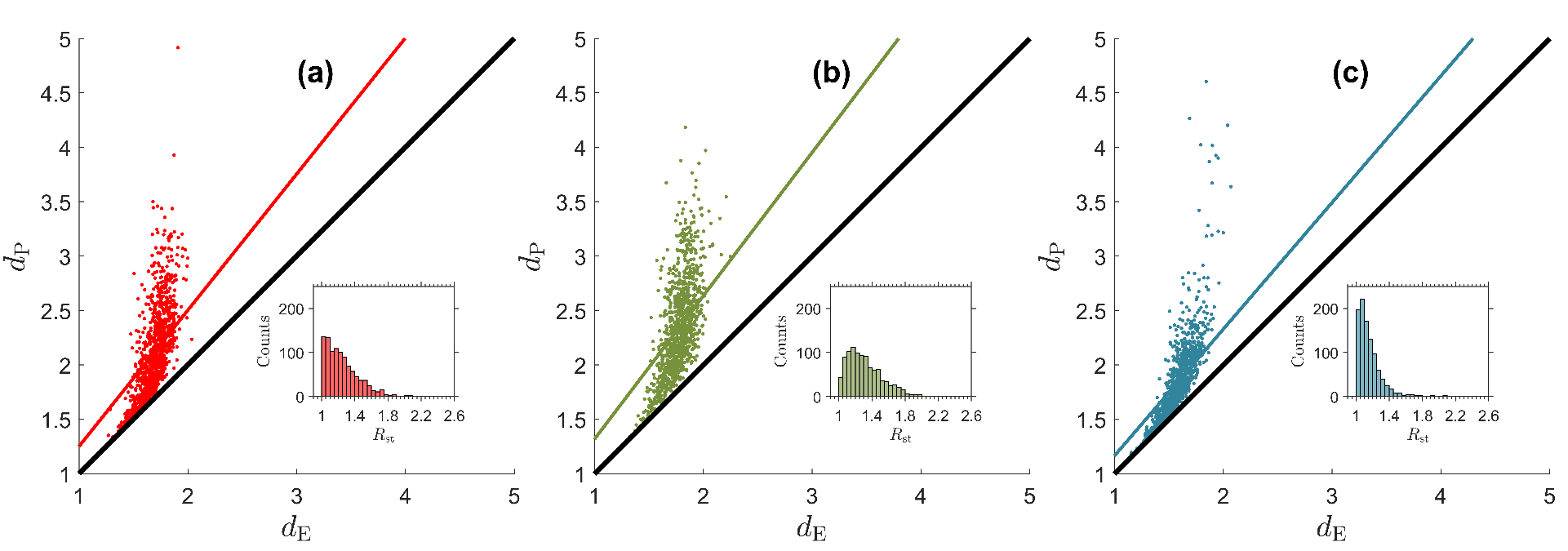}
	\caption{Numerical results of the string method found with the three landscapes. Figures (a)-(c) correspond to the STL, OCL, and UTL. Each subplot illustrates the path length $d_{\rm P}$ versus the Euclidean distance $d_{\rm E}$ constructed to connect one thousand random pairs of optimal controls in the top manifold for each landscape. The inset in each subplot is a histogram for the distribution of the ratio $R_{\rm st}$. The average ratios $R_{\rm st}^{\rm {avg}}$ were 1.2515 for the STL (red line), 1.3171 for the OCL (green line), and 1.1638 for the UTL (blue line), which are close to unity (black lines).}
	\label{fig:EL_vs_PL}
\end{figure*}

Figure~\ref{fig:EL_vs_PL} depicts the correlation of the path length $d_{\rm P}$ of a final string and the Euclidean distance $d_{\rm E}$ between its two ends as a scatter plot over the thousand pairs for each type of the landscape. The minimum, average and maximum values of the ratio $R_{\rm st}$ were denoted by $R^{\min}_{\rm st}$, $R^{\rm {avg}}_{\rm st}$ and $R^{\max}_{\rm st}$, respectively, and are listed in Table~\ref{tab:R_st}. It can be seen that $R_{\rm st}$ values are not significantly larger than $1.0$. The numerical results provide evidence that the top manifold in each landscape of the systems defined by Eqs.~(\ref{eq:four_dim_sys_H0})-(\ref{eq:four_dim_sys_dipole}) is not only path-connected but also appears to have simple features rendering the paths to be modestly curved to avoid those features.

\begin{table}[htbp]
	\centering
	\renewcommand\arraystretch{1.2}
	\caption{Numerical results of the metrics defined in Eq.~(\ref{eq:ratio_R}) reflecting each path-connected curvature with the string method for the three landscapes. One thousand strings were generated between pairs of random controls in each landscape's top manifold.}
	\label{tab:R_st}
	\setlength{\tabcolsep}{5mm}{
		\begin{tabular}{c|c|c|c}
			\hline\hline
			{} & $R_{\rm st}^{\min}$ & $R_{\rm st}^{\rm {avg}}$ & $R_{\rm st}^{\max}$\\
			\hline
			STL & 1.0008 & 1.2515 & 2.5718 \\
			OCL & 1.0038 & 1.3171 & 2.2777 \\
			UTL & 1.0030 & 1.1638 & 2.5229 \\
			\hline
			\hline
	\end{tabular}}
\end{table}

An important matter is whether the height of the near-top manifolds reflected in $J_{\max}-\varepsilon=1-\varepsilon$ affects the path-connectedness findings. To ensure that the tolerance $\varepsilon=0.001$ is sufficient to capture the main character of the top manifold, we generated paths under different values of tolerance  $\varepsilon=10^{-1},10^{-2},10^{-3},10^{-4},10^{-5}$. The simulation results showed little difference among these cases when $\varepsilon\leq 0.01$. Therefore, we fixed $\varepsilon=0.001$ in all the remaining simulations in the paper. We also note that path-connectedness was found even for $\varepsilon=0.1$, suggesting that this property may show up at all landscape level sets of suboptimal value $J<1$. The assessment of this conjecture is left for future study.

We also implemented the D-MORPH connecting algorithm in Sec.~\ref{sec:III}, aiming to connect the pairs of optimal controls in each landscape. These pairs were the same as those employed with the string method. The D-MORPH connecting algorithm almost always succeeded but failed in two cases out of the three thousand pairs; it is not known if the two failed cases out of three thousand tests with the D-MORPH connecting algorithm reflect either numerical artifacts or whether a top manifold feature halted the algorithm. We denote the minimum, average and maximum values of $R_{\rm dm}$ by $R_{\rm dm}^{\min}$, $R_{\rm dm}^{\rm {avg}}$ and $R_{\rm dm}^{\max}$, respectively. Table~\ref{tab_R_dm} reports the statistical results of all successful searches as well as showing the two failed cases. Comparing the results of the string method presented in Table~\ref{tab:R_st} and the D-MORPH connecting method in Table~\ref{tab_R_dm}, the average ratios $R_{\rm st}^{\rm {avg}}$ and $R_{\rm dm}^{\rm {avg}}$ are not only small but also very close to each other, implying the intrinsic ease of movement over the top manifold. Nevertheless, top manifold features are present, as indicated in Figure~\ref{fig:string_convergence}(a), and found in all string tests. It is possible to conceive that irregular shaped features at $J_{\max}-\varepsilon$ could easily halt the D-MORPH path connecting algorithm seeking $E_{\rm start}(t) \rightarrow E_{\rm target}(t)$. However, Table~\ref{tab_R_dm} shows that this behavior rarely happens, suggesting that the D-MORPH path connecting algorithm can easily navigate around the features.

\begin{table}[htbp]
	\centering
	\renewcommand\arraystretch{1.2}
	\caption{Simulation results for the D-MORPH connecting algorithm conducted on the same pairs of optimal control fields used in the string method summarized in Table~\ref{tab:R_st}}.
	\label{tab_R_dm}
	\setlength{\tabcolsep}{3mm}{
		\begin{tabular}{l|c|c|c|c}
			\hline\hline
			{} & $R_{\rm dm}^{\min}$ & $R_{\rm dm}^{\rm {avg}}$ & $R_{\rm dm}^{\max}$ & \makecell[c]{Number of \\failed cases}  \\
			\hline
			STL & 1.0006 & 1.2278 & 2.2257 & 0 \\
			OCL & 1.0028 & 1.3153 & 2.2182 & 1 \\
			UTL & 1.0023 & 1.1590 & 5.7207 & 1 \\
			\hline
			\hline
	\end{tabular}}
\end{table}

\begin{figure}[htbp]
	\centering
	\includegraphics[width=1\columnwidth]{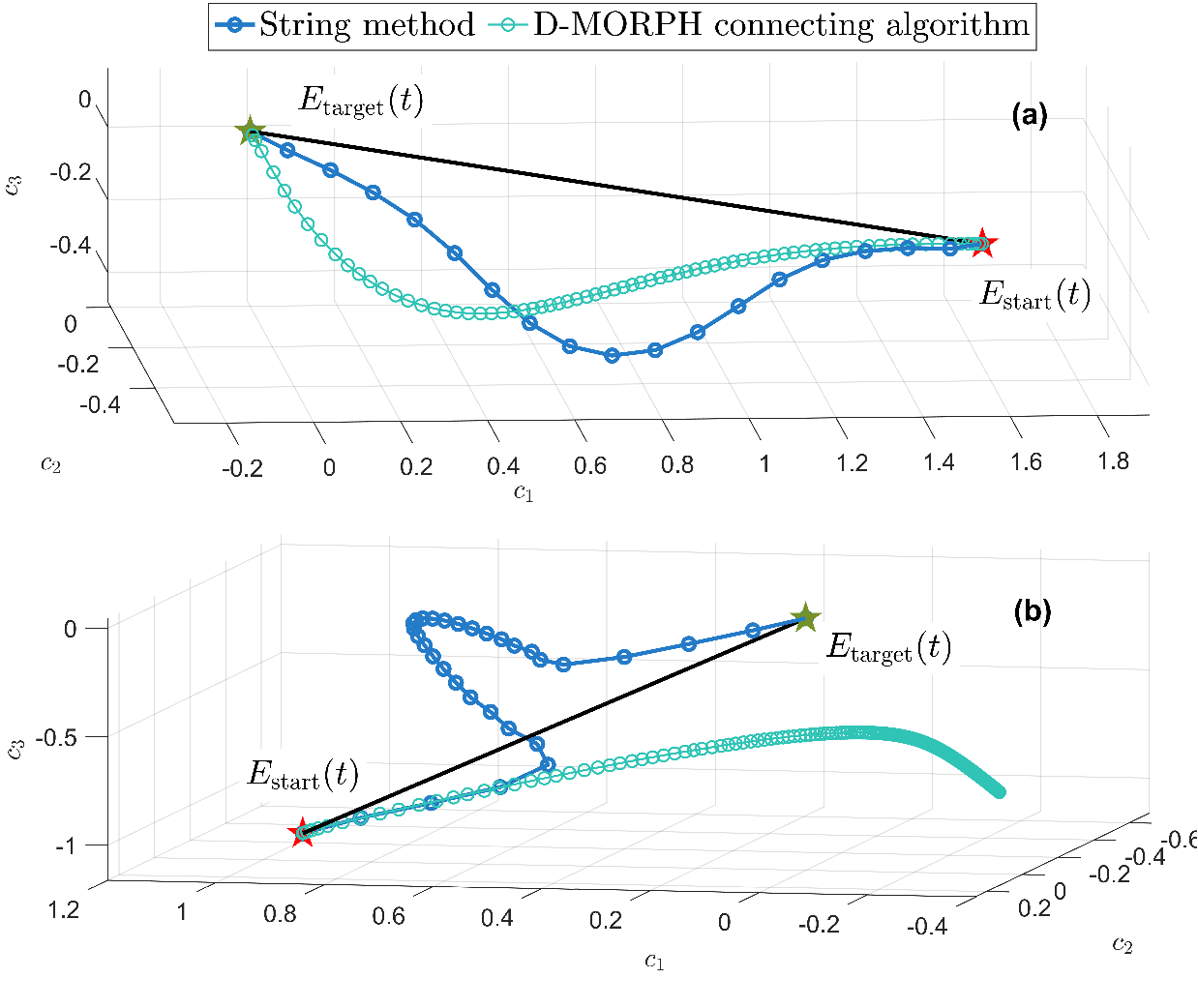}
	\caption{(a) PCA-based visualization of the two illustrative paths, generated by the string method (blue dots) and the D-MORPH connecting algorithm (green dots), that connect the same pair of optimal control fields $E_{\rm {start}}(t)$ (red star) and $E_{\rm {target}}(t)$ (green star). Although each method found a unique path for a pair of fields, the fact that the paths differed shows that path-connectedness between any pair of top manifold fields reflects the global character of the top manifold and the algorithms used to find the connecting paths. The two paths are projected onto the three-dimensional subspace spanned by the basis vectors $\{v_1(t),v_2(t),v_3(t)\}$ solved by PCA. The captured variations were $87.76\%$ for $v_1(t)$, $10.61\%$ for $v_2(t)$, and $1.34\%$ for $v_3(t)$. The total percentage was $99.71\%$. (b) One failure case (i.e, out of the two found in three thousand tests) of the D-MORPH connecting algorithm (green dots), while the string method (blue dots) was always successful.}
	\label{fig:path_comparisons}
\end{figure}

To see the differences between the paths located by the two methods, Figure~\ref{fig:path_comparisons}(a) visualizes the distinctions between the two typical paths connecting the same pair by the PCA algorithm. Figure~\ref{fig:path_comparisons}(b) shows a failure case for the D-MORPH simulation for the UTL case. The D-MORPH connecting algorithm failed to proceed and exhibited a significant deviation along its path. Due to the myopic nature of the D-MORPH search strategy, the exploration chose an improper route and went to a "dead end" in this case. The string method, however, optimized the entire path simultaneously, which may make it be more robust against a broken string due to the presence of local features in the landscape, as explained earlier. Nevertheless, the D-MORPH connecting algorithm almost always connected pairs of the top manifold fields and gave similar $R$ values, and the two rare failures possibly may come from numerical artifacts creating what appears to be an inhibiting top manifold feature.

\section{Additional generality of the Top Manifold connectedness findings}
\label{sec:V}
The results in Sec.~\ref{sec:IV} indicate that the top manifolds of all three landscapes are likely path-connected. However, the simulations were done on a fixed quantum control system, with the fields sampled within a bounded subset. This section further explores the geometry of the top manifold in broader contexts.

\subsection{Path-connectedness tests using disparate optimal controls}
\label{subsec:move_far}
The initial controls in Sec.~\ref{sec:II} were restricted in a bounded region since the trial fields \eqref{eq:initial_fields} were all sampled from a unit sphere (i.e., of fluence $1$) in control space. The resulting set of optimal controls were altered by the gradient flow, whose final fluence values ranged over $[0.98,2.34]$.

\begin{figure}[h]
	\centering
	\includegraphics[width=1\columnwidth]{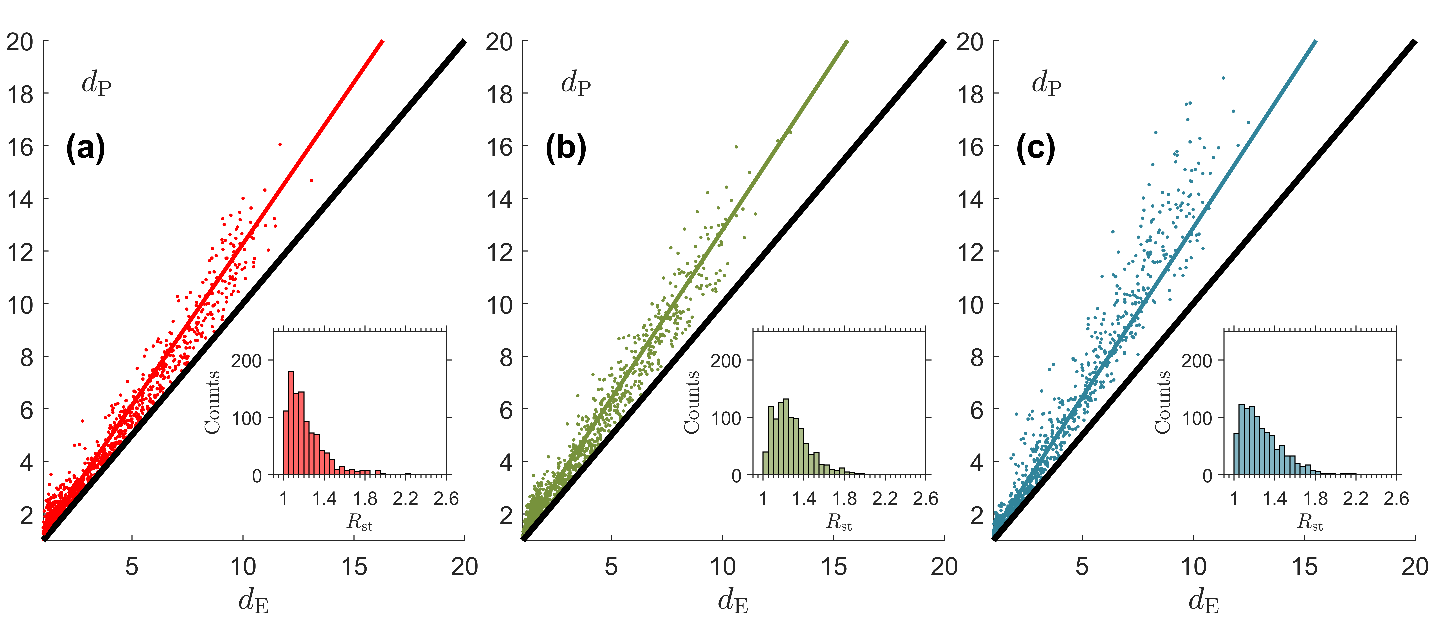}
	\caption{Numerical results of the strings connecting broadly distributed optimal controls in the top manifold. Figures (a)-(c) correspond to STL (red), OCL (green), and UTL (blue), respectively. Each subplot illustrates the path length of a string $d_{\rm P}$ versus its Euclidean distance $d_{\rm E}$. One thousand strings (i.e., pairs of the top manifold fields) were generated for each landscape. The average ratio values $R_{\rm st}^{\rm {avg}}$ of the STL (red line), OCL (green line), and UTL (blue line) are only modestly larger than unity (black lines).}
	\label{fig:el_vs_pl_broad}
\end{figure}

To better understand the prevalence of path-connectedness, here we investigate broader regions of the top manifold. We sampled trial fields with fluence $F_0\in [0.01,100]$, which can be done by setting the variable $A$ in Eq.~(\ref{eq:initial_fields}) so that the $\log_{10}F_0$ obeys the uniform distribution over $\left[-2,2\right]$. One thousand pairs of such optimal controls were generated for each of the three landscapes. All these pairs could be successfully connected via continuous paths in the top manifold solved by the string method. Figure~\ref{fig:el_vs_pl_broad} demonstrates their path lengths as scatter plots, which show that the range of the sampled fields measured by the maximum pairwise distance $d_{\rm E}$ is enlarged by almost four times than that in Figure~\ref{fig:EL_vs_PL}. The average ratio values $R_{\rm st}^{\rm {avg}}$ were 1.227 for the STL, 1.280 for the OCL, and 1.288 for the UTL. Compared with Figure~\ref{fig:EL_vs_PL}, the ratios are evaluated with more widely distributed optimal control fields, yet the connectedness results are similar as evident from comparing with $R_{\rm st}^{\rm {avg}}$ in Table~\ref{tab:R_st}. The results leading to Figure~\ref{fig:el_vs_pl_broad} also indicate the vastness of the top manifold because it includes pairs of fields that are very distant from each other.

The D-MORPH algorithm, employed in a different fashion than before, is also amenable to exploring the spaciousness of the top manifold by asking that the control $E(s,t)$ moves as far as possible from an initial control $E(0,t)$. Similar to the definition in Eq.~(\ref{eq:distance}), let ${\Delta}_{\rm far}(s,t)=E(s,t)-E(0,t)$, and $D_{\rm far}(s)=\|{\Delta}_{\rm far}(s)\|_2^2$, which is the square of the distance from the current field to the initial field. The guiding function in Eq.~(\ref{eq:dmorph_grad}) is accordingly chosen as 
\begin{equation}\label{eq:f_far}
	f_{\rm far}(s,t)=\frac{{\Delta}_{\rm far}(s,t)}{\|{\Delta}_{\rm far}(s)\|_2},~~s>0,
\end{equation}
so that the distance $D_{\rm far}(s)$ from the initial field increases as $s$ grows (see Appendix~\ref{app:guiding_func}). A random initial direction $f_{\rm {far}}(0,t)$ needs to be selected to start the roving. Each walk terminates when the fluence $F=\|E(s)\|_2^2>10^3$ since control fields with very large fluence have wandered rather far. In each landscape, ten optimal controls were generated from the trial fields of the form defined by Eq.~(\ref{eq:initial_fields}), and from each trial field one hundred far-reaching walks were performed from randomly chosen initial directions. All far-reaching walks successfully reached the largest fluence value of $10^3$. The average ratios $R_{\rm {far}}$ of these walks defined by Eq.~(\ref{eq:ratio_R}) were found to be below 1.1, signifying the straightness of these top manifold trajectories. The behavior of these walks affirms that the top manifold is sizable, with generally easy accesses from the optimal controls with small fluence $F\approx 1$ for $E(0,t)$ to those with large fluence $F\approx 10^3$. The wandering behavior towards a large fluence can be understood from the very high dimensional control space, with the outer volume growing rapidly from a weak fluence initial field. Furthermore, the value of $R_{\rm {far}}\lesssim 1.1$ indicates the exterior regions of the top manifold are rather devoid of features.

\subsection{Stochastic exploration in the top manifold}
All of the studies above are based on optimal controls generated in one way or another via tailored gradient flows. Surprisingly, the $R$ values for all explorations were relatively small, statistically implying that the top manifolds are not very gnarled. Here, we will give yet another image via stochastic walks over the top manifold.

Stochastic D-MORPH exploration can be applied in an ergodic fashion by choosing random guiding functions $f(s,t)$~\cite{beltrani2007photonic}. The guiding function $f(s,t)$ was randomly initialized and remained fixed over a window $[(n-1)\Delta s, n\Delta s]$, $n=1,\cdots,n_{\max}$. While transiting through each window, the field takes multiple integration steps following Eq.~(\ref{eq:dmorph_grad}) to explore the top manifold. After each exploration window, the guiding function was randomly reinitialized to steer the trajectory toward another random direction. This process continued until the homotopy variable $s$ reached $s_{\max}=n_{\max}\Delta s$. The randomness embedded in the algorithm adds to its ability to explore potential "hidden" regions in the top manifold.

\begin{figure}[htbp]
	\centering
	\includegraphics[width=1\columnwidth]{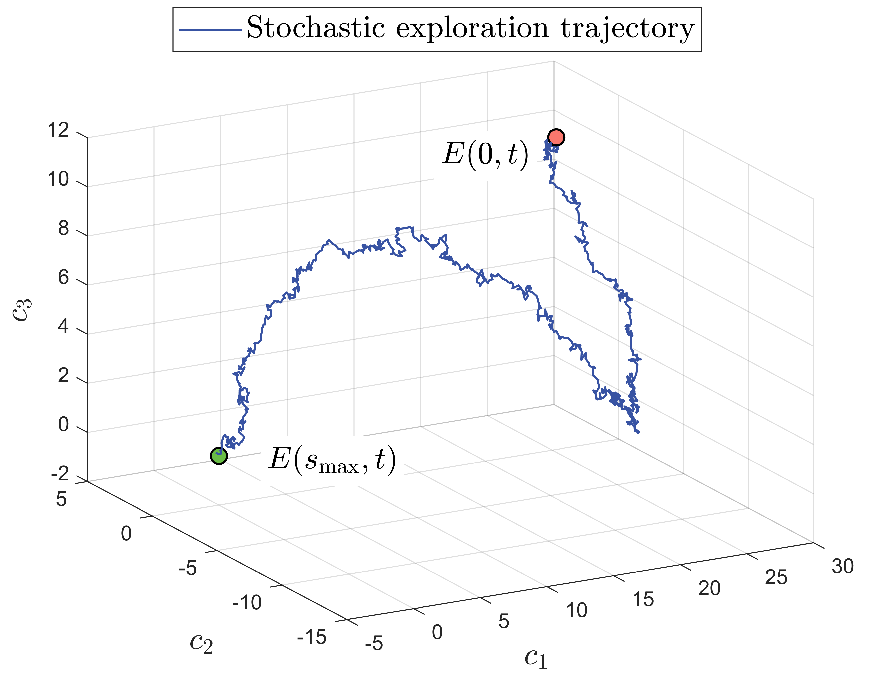}
	\caption{A stochastic exploration in the top manifold for UTL. The exploration (blue line) starts from an optimal control (red dot) and proceeds until reaching $s_{\max}=10^3$ (green dot) steps; no traps to top the algorithm were encountered. The Euclidean distance of the random walk and path length of this trajectory are $d_{\rm E}=30.59$ and $d_{\rm P}=935.40$, respectively. The initial field has the fluence $F=\| E(0,t)\|_2^2=1.43$, while the final field has the fluence $F=\| E(s_{\max},t)\|_2^2=937.21$. The percentages of variation captured by the three basis vectors were $63.38\%$ for $v_1(t)$, $15.04\%$ for $v_2(t)$, and $6.06\%$ for $v_3(t)$, with the total percentage being $84.48\%$.}
	\label{fig:random_walk}
\end{figure}

In each landscape, we randomly chose ten initial optimal controls, from each of which ten stochastic explorations were performed with $\Delta s=1$, and the termination value was set at $n_{\max}=10^3$. None of the trajectories got trapped, implying that the random exploration in the top manifold was free without encountering any "dead ends." Figure~\ref{fig:random_walk} shows one example of the stochastic trajectory for the case of the UTL visualized by the PCA method. The path length $d_{\rm P}^{\Delta s}$ and the ratio $R^{\Delta s}$ of the three hundred trajectories over each period $[s,s+\Delta s]$ were also recorded. It was found that $d^{\Delta s}_{\rm P}\in[0.75,0.99]$ and $R^{\Delta s}<1.1$ for all window periods. Additional statistics for a particular stochastic trajectory are given in the caption of Figure~\ref{fig:random_walk}. The numerical results again confirm that the top manifold has simple features permitting random explorations.

\subsection{The diversity of the top paths}
The numerical test in Sec.~\ref{subsec:IV-B} implements the normal string method, which starts the search from interpolated fields between two optimal controls (see Appendix~\ref{app:string_method}) along a straight line between them. Such an initialization method located a relatively short path in the top manifold, as suggested by the modest $R$ values in Table.~\ref{tab:R_st}. However, the paths in the top manifold that connect two optimal controls are generally not unique (see Figure~\ref{fig:path_comparisons}); this section will further explore the scope of top manifold paths between two optimal controls, using the same Hamiltonian in Eqs.~(\ref{eq:four_dim_sys_H0}) and (\ref{eq:four_dim_sys_dipole}). 

\begin{figure}
	\centering	\includegraphics[width=\columnwidth]{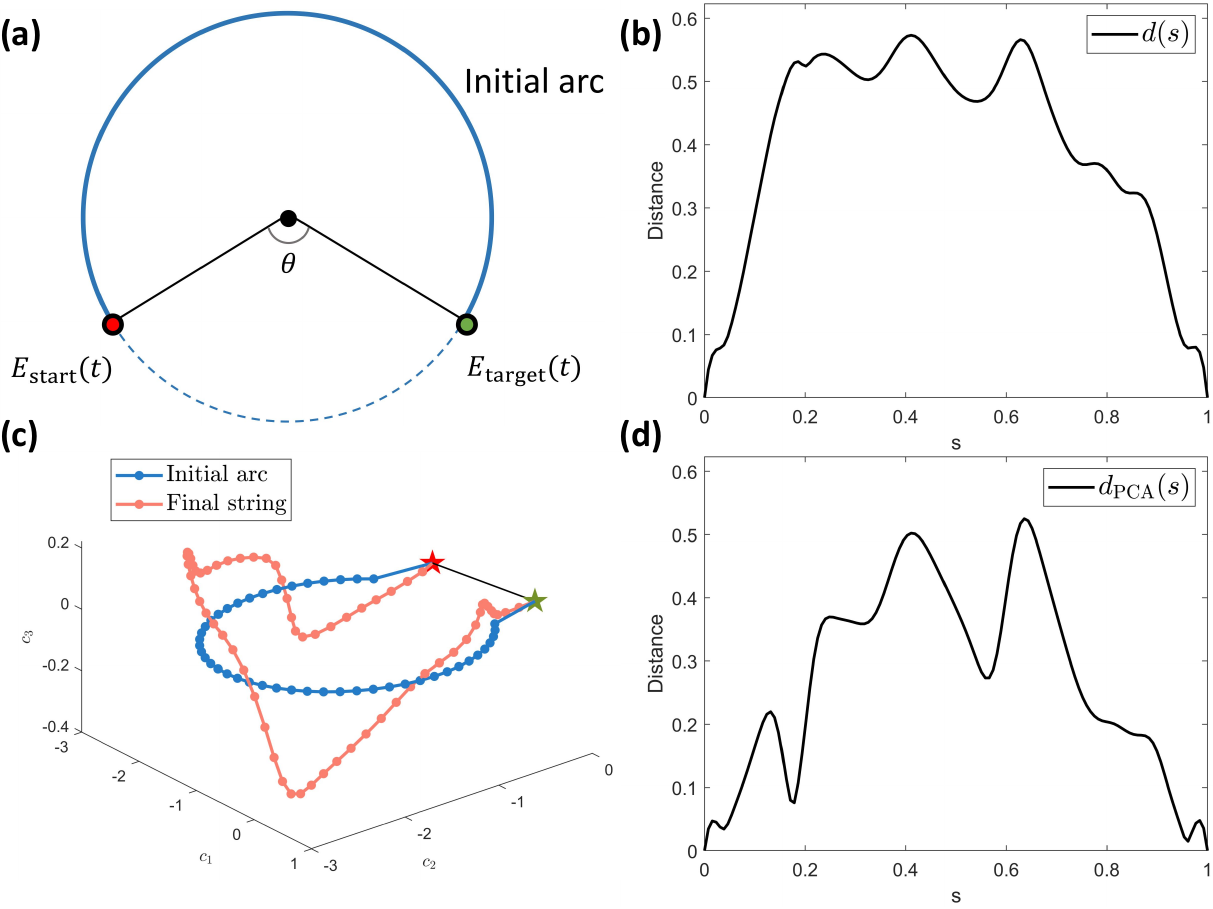}
	\caption{Manifestation of the string method tests starting from arcs in control space. (a) Sketch of the initial arc (blue line), which is cut out by the pair of optimal fields (red and green dots), depicted in a two-dimensional plane. The initial $R_{\rm st}$ value defined by Eq.~\eqref{eq:ratio_R} is fully determined by the central angle $\theta$. We conducted numerical tests for $R_{\rm st}=5$ and $10$, whose corresponding control angles in radians are $\theta = 1.0917$ and $0.5785$, respectively. Figures (b)-(d) show the results of one representative case in the STL scenario. (b) Distance $d(s)$ between the initial arc and the final string. (c) PCA-based visualization of the initial arc (blue) and the final string (red). (d) Distance $d_{\rm PCA}(s)$ between the two low-dimensional projected paths shown in (c).} 
	\label{fig:arc_test}
\end{figure}

To investigate the diversity of the top paths as well as test the exploratory capability of the string method, we purposely made the initial strings rather curved. We randomly sampled one hundred pairs of optimal controls in the top manifold. For each pair, we designed the arc of a circle to lie in a random plane in the high-dimensional control space that passed through the two fields. Figure~\ref{fig:arc_test}(a) sketches such an arc in the plane. The extent of the arc cut out by the pair of fields can be arbitrarily made by choosing the central angle $\theta$. Therefore, we tailored the one hundred arcs in the random planes to have $R_{\rm st}=5$ (i.e., $R_{\rm st}$ here is the length along the arc divided by the Euclidean distance between $E_{\rm start}(t)$ and $E_{\rm target}(t)$) and set them as the initial strings. The string method successfully transformed all arcs into strings with $N_{\rm st}\in [32, 92]$ intermediate fields in the top manifold, whose average $R_{\rm st}$ values were 5.539 for the STL, 5.809 for the OCL, and 5.795 for the UTL. These values were close to 5, implying that the initial arcs were not dramatically deformed, as will be verified below.
Here we show one STL case in the  above scenario as an example. We parameterized the initial arc and the final string as $E_{\rm arc}(s,t)$ and $E_{\rm final}(s,t)$, in which $s\in[0,1]$ marks the relative position of the control field along the path with $E_{\rm start}(t)=E_{\rm final}(0,t)=E_{\rm arc}(0,t)$ and $E_{\rm target}(t)=E_{\rm arc}(1,t)=E_{\rm final}(1,t)$. The distance between the two paths were measured by $d(s):=\|E_{\rm arc}(s,t)-E_{\rm final}(s,t)\|_2$, which is plotted in Figure~\ref{fig:arc_test}(b). The distance $d(s)$ is smaller than $0.60$ for all values of $s$, implying that the two paths are close in control space. The PCA-based visualization of the initial arc and the final string is shown in Figure~\ref{fig:arc_test}(c), and the distance $d_{\rm PCA}(s)$ between the two projected three-dimensional paths is plotted in Figure~\ref{fig:arc_test}(d). We see that $d_{\rm PCA}(s)$ is also smaller than $0.60$ and resembles the exact distance $d(s)$. The mean distance, defined by $\bar{d}=\int_0^1 d(s)\: ds$, was determined between every initial arc and its corresponding final morphed string, whose average values for the hundreds of pairs were 0.350 for the STL, 0.406 for the OCL, and 0.584 for the UTL. These numerical outcomes indicate that, even when the initial string is far from straight, the string method always locates a top path close to the initial string. We also tested the initial arcs with $R_{\rm st}=10$, and the observations were similar. All these results collectively suggest that top manifold connecting paths are diverse and prevalent in control space.

\subsection{Path-connectedness tests with additional classes of quantum systems}
\label{subsec:add_sys}
The simulations in the previous sections provide strong numerical evidence suggesting path-connectedness of the top manifold based on working with a controllable four-level quantum system with a single control field. To verify the generality of the results, we investigate several model variations, including differing dipole moment matrices, system dimensions, and Hamiltonian structures.

We first assessed a five-dimensional system with the following Hamiltonian:
\begin{equation}\label{eq:five_dim_sys}
	\begin{aligned}
		H_0&=\begin{pmatrix} 
			-8 & 0 & 0 & 0 & 0 \\
			0 & -6 & 0 & 0 & 0 \\ 
			0 & 0 & -2 & 0 & 0 \\
			0 & 0 &  0 & 4 & 0 \\
			0 & 0 &  0 & 0 & 12
		\end{pmatrix},\\
		\mu&=\begin{pmatrix} 
			0 & \pm1 & 0 & 0 & 0 \\ 
			\pm1 & 0 & \pm1 & 0 & 0 \\ 
			0 & \pm1 & 0 & \pm1 & 0 \\ 
			0 & 0 & \pm1 & 0 & \pm1 \\
			0 & 0 & 0 & \pm1 & 0 \\
		\end{pmatrix}.
	\end{aligned}
\end{equation}
The random signs were chosen to keep $\mu$ real symmetric. The dipole moment $\mu$ only permits transitions between adjacent states. For the STL, $|i\rangle$=$|1\rangle$ and $|f\rangle=|5\rangle$. For the UTL, $\rho=\rm {diag}\{0.5,0.5,0,0,0\}$ and $\theta=\rm {diag}\{0,0,0,1,1\}$. For the UTL, a quasi-random Hermitian matrix $G$ was generated with zero trace, and the target unitary transformation was set as $W=\exp(iG)$~\cite{riviello2015searching}. To cover all allowed transition frequencies, the components $\{\omega_m:m=1,\cdots,100\}$ of the initial trial fields defined by Eq.~(\ref{eq:initial_fields}) were drawn from the uniform distribution over $[1,9]$. Other parameter configurations remained the same as those in Sec.~\ref{sec:IV}.

The second system we considered was a coupled two-spin system with multiple control fields~\cite{sun2015experimental}. The system Hamiltonian had the form
\begin{equation}\label{eq:multi_control_sys}
	H(t)=2\pi J_{IS}I_zS_z+u_x^I(t) I_x+u_y^I(t)I_y+u^S_x(t)S_x+u^S_y(t)S_y,
\end{equation}
where $I_j=\frac{\sigma_j}{2}\otimes  \mathbb{I}_2$, $S_j=\mathbb{I}_2\otimes \frac{\sigma_j}{2}$, $\sigma_j\in\{\sigma_x,\sigma_y,\sigma_z\}$ are Pauli matrices, $\mathbb{I}_2$ is the two-dimensional identity matrix, and $J_{IS}$ is the coupling strength. $\{u_x^I(t),u_y^I(t),u^S_x(t),u^S_y(t)\}$ are four controls locally imposed on the two spins. They are concatenated to form the complete control $\tilde{u}(t)=[u_x^I(t),u_y^I(t),u^S_x(t),u^S_y(t)]$. $J_{IS}=1$ was specified in the simulations. Each control field was independently initialized as the form in Eq.~(\ref{eq:initial_fields}). The control pulse duration was $T=20$ and the frequency interval was $\omega_m\in [2.5,3.5]$. The other parameters and the computational top manifold tests were the same as those in Sec.~\ref{sec:IV}, based on the string method.

Similar simulations were conducted on the above two systems. One thousand optimal controls were generated in each system, from which one thousand pairs were randomly selected for connectedness tests by the string method. The gradient ascent algorithm was easily extended to the multiple-control system in Eq.~(\ref{eq:multi_control_sys})~\cite{sun2015experimental}. All strings successfully converged to well defined paths in the top manifold. The average ratios $R_{\rm st}^{\rm {avg}}$ of the two systems recorded in Table~\ref{tab:additional} are modestly larger than unity, so these paths are still relatively straight.

\begin{table}[htbp]
	\centering
	\renewcommand\arraystretch{1.2}
	\caption{The average ratios $R_{\rm st}^{\rm {avg}}$ of the converged strings of the two additional systems in Sec.~\ref{subsec:add_sys}. One thousand strings were generated in each landscape. In the table, Systems~(1) and~(2), respectively, refer to the five-dimensional case in Eq.~(\ref{eq:five_dim_sys}) and the multiple-control case in Eq.~(\ref{eq:multi_control_sys}).}
	\label{tab:additional}
	\setlength{\tabcolsep}{5mm}{ 
		\begin{tabular}{l|c|c}
			\hline\hline
			{} & System~(1) & System~(2)\\
			\hline
			STL & 1.3409 & 1.2256 \\
			OCL & 1.3929 & 1.2764 \\
			UTL & 1.3750 & 1.2699 \\
			\hline
			\hline
	\end{tabular}}
\end{table}

\section{Conclusions}
\label{sec:VI}
This paper mainly used the string method and the D-MORPH algorithm to explore path-connectedness of the top manifold of QCLs by seeking continuous paths (i.e., expressed as a set of uniformly spaced fields) between randomly sampled optimal solutions. From several quantum control systems whose control resources are not limited, the simulation results collectively show that the top manifold appears to be path-connected in the studied landscapes for state transitions, observable control, and unitary transformations. Moreover, although the paths connecting optimal solutions are not straight, most are only mildly curved, with the average straightness ratios $R$ close to unity. Nevertheless, the top manifold does have distributed features characterized by the objective value $J$ with initial fields along the string satisfying $J<1$. The overwhelming success of the D-MORPH path connecting algorithm implies that these top landscape features are easy to navigate around. Top manifold features were seen in a prior limited study~\cite[Figure~9]{tibbetts2015constrained}, and the present work provides extensive evidence for their presence. Constrained controls in any application can lead to such features, but that is not believed to be the case in the present work. Further study is needed to fully understand the physical and mathematical origin and nature of features in the top manifold.

From the collective numerical findings, we conjecture that the top manifold is path-connected if the quantum system satisfies conditions (i)-(iii) in Sec.~\ref{subsec:II-A}. This paper provides a foundation for further rigorous analysis of the top manifold connectedness. The conclusions of the paper and the methods developed for locating paths in the top manifold are efficient, and thus offer various algorithms for seeking optimal control fields that meet desirable ancillary goals such as higher robustness or other enhanced properties of the quantum system under control. The small $R$ values of the paths in the top manifold are reminiscent of previous findings about the near straightness of the gradient flows for finding optimal controls~\cite{nanduri2013exploring}, which also implies an overall simple structure of the quantum control landscape. It is an open question whether this common behavior (i.e., while climbing and moving in the top manifold) is related. These topics will be explored in future studies.

\appendix
\section{Details of the string method procedures}
\label{app:string_method}
The string method expressed initially as a straight path is executed as follows: 
\begin{enumerate}[(I)]
	\item A straight line of $N_{\rm st}+1$ equally spaced interpolated fields $\{\hat{E}_i(t):\hat{E}_i(t)=(1-\frac{i}{N_{\rm st}})\hat{E}_0(t)+\frac{i}{N_{\rm st}}\hat{E}_{N_{\rm st}}(t),\:  i=0,\cdots, N_{\rm st}\}$ (referred to as images in Ref.~\cite{weinan2002string}) is initialized in control space, of which the two ends are fixed at a given pair of optimal fields, i.e., $\hat{E}_{0}(t)=E_{\rm {start}}(t)$ and $\hat{E}_{N_{\rm st}}(t)=E_{\rm {target}}(t)$. 
	\item The value of $J[E_i(t)]$ is evaluated, and for generality here, we assume $J[E_i(t)]<1-\varepsilon$. Thus, each interpolated field $\{\hat{E}_i(t): i=1,\cdots,N_{\rm st}-1\}$ takes one gradient ascent step according to Eq.~(\ref{eq:gradient_flow}). The fixed step fourth-order Runge-Kutta method is employed to solve Eq.~(\ref{eq:gradient_flow}). The sampled string is lifted in the landscape as the interpolated fields are pushed towards the top.
	\item Redistribute the interpolated fields $\{\hat{E}_i(t):i=1,\cdots,N_{\rm st}-1\}$ uniformly along the string. To prevent the interpolated fields from converging directly towards either of the two endpoints, the string method imposes the constraint that the distances between adjacent fields are equal. This constraint leads to the redistribution of the interpolated fields along the sampled string, which was achieved by cubic spline interpolation~\cite{weinan2007simplified}.
	\item Insert one additional field to improve the resolution if the gap between the sampled string and the landscape is out of tolerance (as explained below). A string sampled by the fields $\{\hat{E}_i(t):i=0,1,\cdots,N_{\rm st}\}$ is piece-wise straight in control space. The interpolated control objective value $J$ at the position $\alpha$ between adjacent fields $\hat{E}_i(t)$ and $\hat{E}_{i+1}(t)$ is given by linear interpolation
	\begin{equation}
		J_{\rm int}^{i,i+1}(\alpha)=(1-\alpha) J[\hat{E}_i(t)]+\alpha J[\hat{E}_{i+1}(t)],
	\end{equation} 
	where $\alpha\in [0,1]$. The true objective value at $\alpha$ is 
	\begin{equation}\label{eq:string_interpolate}
		J^{i,i+1}(\alpha)=J[(1-\alpha) \hat{E}_i(t)+\alpha \hat{E}_{i+1}(t)].	
	\end{equation}
	If the gap between the interpolated value and the true value is above a given tolerance $\varepsilon_{\rm st}$, i.e.,
	\begin{equation}\label{eq:string_above_thre}
		J_{\rm int}^{i,i+1}(\alpha)-J^{i,i+1}(\alpha)>\varepsilon_{\rm st},
	\end{equation}
	then one additional field is inserted at the position $\alpha$ between $\hat{E}_{i}(t)$ and $\hat{E}_{i+1}(t)$, and $N_{\rm st}~\rightarrow~ N_{\rm st}+1$, accordingly. In the simulations, $\alpha$ is discretized into a finite set of grid points. Moreover, only one field is inserted in one iteration at the position where the gap is the largest.  After the insertion, the interpolated fields are redistributed following step (III). 
	
	\item Repeat steps (II)-(IV) until the string satisfies either of the following termination conditions: (i) Successful convergence: All control fields along the string are identified in the top manifold, and the resolution of the string is qualified according to step (IV). (ii) Failure: The number of iterations is taken out to the allowed maximum (1000 in our experiments) while the string has not yet converged. 
\end{enumerate}
The string method corresponding to steps (I)-(III) provides the update rule and the constraints upon the fields. Inspired by AutoNEB, step (IV) increases the resolution of the string when necessary for better approximation. As a heuristic algorithm, the string method and its modified forms have been successfully employed to study the connectedness of loss landscapes in a wide range of areas and exhibited excellent performance~\cite{draxler2018essentially,hamilton2022mode,sheppard2008optimization}.

\section{The significance of the guiding functions \texorpdfstring{$f_{\rm {dist}}(s,t)$}{Lg} and \texorpdfstring{$f_{\rm {far}}(s,t)$}{Lg}}
\label{app:guiding_func}
The distance $D(s)$ from the initial field to the target field, defined in Eq.~(\ref{eq:distance}), changes under the guiding function $f_{\rm {dist}}(s,t)$ in Eq.~(\ref{eq:guiding_func}) as
\begin{equation}\label{eq:B1}
	\begin{aligned}
		\frac{d D(s)}{d s}&=2\int_0^T \Delta^T(s,t)\frac{\partial \Delta(s,t)}{\partial s}\: dt \\
		& = -2\int_0^T \Delta^T(s,t)[\mathbf{1}-\hat{P}]\cdot f_{\rm {dist}}(s,t)\: dt\\
		& = -\frac{2}{\|\Delta(s)\|_2}\int_0^T\Delta^T(s,t)[\mathbf{1}-\hat{P}]\cdot \Delta(s,t)\: dt.
	\end{aligned}
\end{equation}
where $f_{\rm {dist}}(s,t):=\frac{\Delta(s,t)}{\|\Delta(s)\|_2}$ in Eq.~(\ref{eq:guiding_func}). Since $\mathbf{1}-P$ is a projection operator that is positive semi-definite, we can see that $\frac{d D(s)}{ds}\leq 0$ according to Eq.~\eqref{eq:B1}.


Similar to the above derivation, the distance $D_{\rm far}(s)$ defined in Sec.~\ref{subsec:move_far} changes under the guiding function $f_{\rm far}(s,t)$ in Eq.~(\ref{eq:f_far}) as
\begin{equation} 
	\frac{dD_{\rm far}(s)}{ds}=\frac{2}{\|\Delta_{\rm far}(s)\|_2}\int_0^T\Delta_{\rm far}^T(s,t)[\mathbf{1}-\hat{P}]\cdot \Delta_{\rm far}(s,t)\: dt\geq 0. 
\end{equation} 
This exploration is designed to move further from the initial field, as done in Sec.~\ref{subsec:move_far}.

\section{D-MORPH connecting algorithm}
\label{app:dmorph_connecting}
The procedure of the D-MORPH connecting algorithm is the following:
\begin{enumerate}[(I)]
	\item Given a pair of optimal fields $E_{\rm {start}}(t)$ and $E_{\rm {target}}(t)$, set  $E(0,t)=E_{\rm {start}}(t)$.
	\item Explore the top manifold by solving Eq.~(\ref{eq:dmorph_grad}) with $f(s,t)=f_{\rm {dist}}(s,t)$. The exploration terminates if the trajectory "falls off," i.e.,  $J\left[E(s,t)\right]<J\left[E(0,t)\right]-\varepsilon_{\rm dm}$, where $\varepsilon_{\rm dm}$ is a small tolerance. Here, "falling" from the top is ascribed to numerical inaccuracies.
	\item Implement the gradient ascent in Eq.~(\ref{eq:gradient_flow}) to push the trajectory back to the top manifold.
	\item Repeat steps (II)-(III) until one of the following conditions holds: (i) $\|\frac{\partial E(s,t)}{\partial s}\|_2<\tau_1$, or (ii) $\|E(s,t)-E_{\rm {target}}(t)\|_2<\tau_2$, where $\tau_1$ and $\tau_2$ are small thresholds.
\end{enumerate}
In step (IV), condition (i) means the search trajectory is trapped and exhibits difficulty in moving. In contrast, the exploration is considered to reach (the vicinity of) the target field $E_{\rm {target}}(t)$ if condition (ii) is satisfied.

\section{PCA-based visualization}
\label{app:pca}
Given a pair of optimal controls $\{E_{\rm start}(t), E_{\rm target}(t)\}$, we can determine one or more paths in the top manifold that connect them. PCA method can project all these paths onto a shared, informative low-dimensional space for visualization. The procedures are the following:
\begin{enumerate}[(I)]
    \item Let $\{\hat{E}_1(t),\cdots,\hat{E}_N(t)\}$ denote all interpolated fields that sample one or more paths in the top manifold connecting $E_{\rm start}(t)$ and $E_{\rm target}(t)$. Compute the matrix 
	\begin{equation}
		\begin{aligned}
			M=[&\hat{E}_{0}(t)-\hat{E}_{N+1}(t),\hat{E}_{1}(t)-\hat{E}_{N+1}(t),\cdots, \\
			&\hat{E}_{N}(t)-\hat{E}_{N+1}(t)],
		\end{aligned}
	\end{equation} 
    where $\hat{E}_0(t):=E_{\rm start}(t)$ and $\hat{E}_{N+1}(t):=E_{\rm target}(t)$.
    
    \item Apply PCA to the matrix $M$, and select the three most explanatory orthonormal basis vectors $\{v_1(t),v_2(t),v_3(t)\}$. This step can be conducted by a MATLAB routine \verb+pca+, which outputs the required basis vectors and gives the percentage of variance explained by each vector.

    \item Project the string onto the subspace spanned by $\{v_1(t),v_2(t),v_3(t)\}$. To be specific, the three-dimensional coordinate image of $\hat{E}_i(t)$ is $c_n^i=\int_0^Tv_n(t)\hat{E}_i(t) dt$, $n=1,2,3$. One can visualize the fields by plotting the coordinates $(c_1^i,c_2^i,c_3^i)$, $i=0,\cdots,N+1$ in the three-dimensional space.
\end{enumerate}

\textbf{Acknowledgements}: The author RBW acknowledges support from Innovation Program for Quantum Science and
Technology (No.2021ZDXX) and NSFC grant 62173201. The authors H.R., T.-S. Ho, and G. Bhole acknowledge support from the U.S. Department of Energy grant (DE-FG02-02ER15344). 

\bibliography{apssamp}
\end{document}